%% file: main.tex
\documentclass[conference,letterpaper,10pt]{IEEEtran}
\IEEEoverridecommandlockouts
\input{packages.tex}
\input{macros.tex}
\begin{document}
%
%
\title{Threshold Selection for Iterative Decoding\\ of $(v,w)$-regular Binary Codes\\ - extended version -}
%
%
\author{%
\IEEEauthorblockN{Alessandro~Annechini and Alessandro~Barenghi and
Gerardo~Pelosi~\IEEEmembership{Member,~IEEE}}
\IEEEauthorblockA{%
Department of Electronics, Information and Bioengineering - DEIB\\
Politecnico di Milano, Milano, Italy\\
Email: alessandro.annechini@mail.polimi.it, alessandro.barenghi@polimi.it,
gerardo.pelosi@polimi.it}
}
%
%
\maketitle
%
%
\begin{abstract}
\input{abstract.tex}

\end{abstract}
%
%
\section{Introduction}\label{sec:intro}
\input{intro.tex}
%
%
\section{Preliminaries}\label{sec:background}
\input{background.tex}
%
%
\section{Distribution of the Syndrome Weight \texorpdfstring{$\weight{s^{(1)}}$}{}}
\label{sec:syn2it}
\input{syndrome_weight_distribution.tex}
%
%
\section{Second iteration DFR and Thresholds}\label{sec:model}
\input{dfrmodel.tex}
%
%
\section{Numerical validation}
\input{expres.tex}
%
%
%
%
\section*{Acknowledgements} This work was carried out with partial financial
support of the Italian MUR PRIN $2022$ project POINTER ID-$2022$M$2$JLF$2$ and
project SERICS (PE$00000014$) under the NRRP MUR program funded by the EU - NGEU.
%
%
\balance
\bibliographystyle{IEEEtran}
\bibliography{biblio.bib}
%
%
\clearpage
\appendices
\section{Transition probabilities of the non homogeneous Markov process}\label{app:trmatrix}
\input{appendix0}
%
%
\section{Second iteration bit flipping probabilities}\label{app:flipprob}
\input{appendix2}
%
%
\section{Joint distribution of $(\mathcal{F},\mathcal{F}_{01},\mathcal{F}_{11})$}\label{app:distrf}
\input{appendix1}
%
%
\section{DFR estimation with regularity constraints}\label{app:regulardfr}
\input{appendix3}
%
%
\section{Additional numerical results}\label{app:additional_results}
\input{appendix4}

%
%
\end{document}

%% file: packages.tex
\usepackage[utf8]{inputenc}
\usepackage[T1]{fontenc}
\usepackage[caption=false,font=footnotesize,labelfont=sf,textfont=sf]{subfig}
\usepackage{array}
\usepackage[unicode]{hyperref}
\usepackage{url}
\usepackage{balance}
\usepackage{ifthen}
\usepackage{cite}
\usepackage[cmex10]{amsmath} 
\usepackage{amsfonts}
\usepackage{booktabs}
\usepackage[ruled,noline,linesnumbered,noend]{algorithm2e}
\usepackage{multirow}
\usepackage[pdftex]{graphicx}
\graphicspath{{./figures/}}
\DeclareGraphicsExtensions{.pdf,.jpeg,.png}
\usepackage{tikz}
\usepackage{pgfplots}
\pgfplotsset{compat=1.18}
\definecolor{darkred}{RGB}{202,0,32}
\definecolor{darkblue}{RGB}{5,113,176}
\definecolor{lightred}{RGB}{244,165,130}
\definecolor{lightblue}{RGB}{146,197,222}
\pgfplotscreateplotcyclelist{waterfall-floor}{%
red,mark=.\\%
red,only marks, mark=x\\%
orange,mark=.\\%
orange,only marks, mark=x\\%
green,mark=.\\%
green,only marks, mark=x\\%
blue,mark=.\\%
blue,only marks, mark=x\\%
violet,mark=.\\%
violet,only marks, mark=x\\%
}

\pgfplotscreateplotcyclelist{waterfall-floor-mc}{%
red,only marks, mark=x\\%
orange,only marks, mark=x\\%
green,only marks, mark=x\\%
blue,only marks, mark=x\\%
violet,only marks, mark=x\\%
}

\pgfplotscreateplotcyclelist{waterfall-floor-model}{%
red,mark=.\\%
red,dashed, mark=.\\%
orange,mark=.\\%
orange,dashed, mark=.\\%
green,mark=.\\%
green,dashed, mark=.\\%
blue,mark=.\\%
blue,dashed, mark=.\\%
violet,mark=.\\%
violet,dashed, mark=.\\%
}

%% file: macros.tex
\newcommand{\weight}[1]{\ensuremath{\mathtt{wt}\left(#1\right)}}

\newcommand{\sevent}[1]{\ensuremath{\mathtt{E}_{(#1)}}}

\newcommand{\prob}[1]{\ensuremath{\Pr(#1)}}
\newcommand{\condprob}[2]{\ensuremath{\Pr\left(#1\, \mid \, #2\right)}}

\newcommand{\punsatz}{\ensuremath{p_{\mathtt{unsat}|0}}}
\newcommand{\punsato}{\ensuremath{p_{\mathtt{unsat}|1}}}
\newcommand{\upc}{\ensuremath{\mathtt{upc}_j}}

\newcommand{\discre}{\ensuremath{\mathrm{d}}}
\newcommand{\discplus}{\ensuremath{\mathrm{d}_+}}
\newcommand{\discminus}{\ensuremath{\mathrm{d}_-}}

\newcommand{\deltapy}[1]{\ensuremath{\delta_{+,y}(#1)}}
\newcommand{\deltamy}[1]{\ensuremath{\delta_{-,y}(#1)}}

\newcommand{\jset}[2]{\ensuremath{\mathbf{J}_{#1,#2}}}
\newcommand{\bindist}[3]{\ensuremath{\text{\sc Bin}(#1,#2,#3)}}

\newcommand{\pflipzstd}{\ensuremath{p_{\mathtt{flip}|0}}}
\newcommand{\pnoflipostd}{\ensuremath{p_{\neg\mathtt{flip}|1}}}

\newcommand{\pflipz}[1]{\ensuremath{p_{\mathtt{flip}|0,#1}}}
\newcommand{\pnoflipo}[1]{\ensuremath{p_{\neg\mathtt{flip}|1,#1}}}

\newcommand{\pflipzz}{\ensuremath{p_{\mathtt{flip}|00}}}
\newcommand{\pflipzo}{\ensuremath{p_{\mathtt{flip}|01}}}
\newcommand{\pflipoz}{\ensuremath{p_{\mathtt{flip}|10}}}
\newcommand{\pflipoo}{\ensuremath{p_{\mathtt{flip}|11}}}

\newcommand{\synwsit}[2]{\ensuremath{\mathcal{W}^{(1)}_{#1,#2}}}
\newcommand{\discvector}{\ensuremath{\Bar{d}^{(1)}}}
\newcommand{\discvectoridx}[1]{\ensuremath{\Bar{d}^{(1)}_{#1}}}

\newcommand{\thfirstit}{\ensuremath{\mathtt{th}^{(1)}(y)}}
\newcommand{\thsecondit}{\ensuremath{\mathtt{th}^{(2)}(y,z_0,z_1)}}
\newcommand{\optthfirstit}{\ensuremath{\widehat{\mathtt{th}}^{(1)}(y)}}
\newcommand{\optthsecondit}{\ensuremath{\widehat{\mathtt{th}}^{(2)}(y,z_0,z_1)}}

\newcommand{\optthseconditsimple}{\ensuremath{\widehat{\mathtt{th}}^{(2)}(z_0,z_1)}}

\newcommand{\upcfirstit}[1]{\ensuremath{\mathtt{upc}^{(1)}_{#1}}}
\newcommand{\upcsecondit}[1]{\ensuremath{\mathtt{upc}^{(2)}_{#1}}}

\newcommand{\pflipsyn}[2]{\ensuremath{\pi_{flip|(#1,#2)}}}

\newcommand{\pcheck}[3]{\ensuremath{p_{#1,#2 \to #3}}}

\newcommand{\mathalert}[1]{\ensuremath{{\color{blue}#1}}}

\DeclareMathOperator*{\argmin}{arg\,min}

%% file: abstract.tex
THIS PAPER IS ELIGIBLE FOR THE STUDENT PAPER AWARD --
Iterative bit flipping decoders are an efficient and effective decoder choice
for decoding codes which admit a sparse parity-check matrix.
Among these, sparse $(v,w)$-regular codes, which include LDPC and MDPC codes are
of particular interest both for efficient data correction and the design of
cryptographic primitives.
In attaining the decoding the choice of the bit flipping thresholds, which
can be determined either statically, or during the decoder execution by using
information coming from the initial syndrome value and its updates.
In this work, we analyze a two-iterations parallel hard decision bit flipping
decoders and propose concrete criteria for threshold determination, backed by
a closed form model.
In doing so, we introduce a new tightly fitting model for the distribution of the
Hamming weight of the syndrome after the first decoder iteration and
substantial improvements on the DFR estimation with respect to existing approaches.

%% file: intro.tex
\noindent
This work focuses on sparse binary codes with block length $n$, dimension $k$,
and redundancy $r$$=$$n$$-$$k$ that can be described by a $r$$\times$$n$ parity-check
matrix $H = [h_{i,j}]$, with $i$$\in$$\{0,\ldots,r$$-$$1\}$, $j$$\in$
$\{0, \ldots, n$$-$$1\}$, having constant numbers of asserted bits in every
row and column, which in turn are referred to as row- and column-weight and
denoted as $w$ and $v$, respectively.
Such codes are referred to as $(v, w)$-regular codes satisfying,
by construction, the constraint $\frac{n}{r} = \frac{w}{v}$, and consequently
exibiting a rate $\frac{k}{n} = 1-\frac{v}{w}$.
Among the binary Low Density Parity Check (LDPC)~\cite{DBLP:journals/tit/Gallager62,DBLP:journals/tit/LitsynS02}
and the binary Moderate Density Parity Check (MDPC)~\cite{
DBLP:journals/corr/abs-0911-3262, DBLP:conf/isit/MisoczkiTSB13} codes, there are
$(v, w)$-regular codes that admit a sparse parity-check matrix amenable
to efficient linear-time decoding algorithms, with column and row weights
in the range of $\mathcal{O}(\log(n))$ for LDPCs and
$\mathcal{O}(\sqrt{n\log(n)})$ for MDPCs.
As a consequence, sparse regular codes are currently among the most interesting
ones for engineering applications concerning error correction and
post-quantum cryptography~\cite{DBLP:journals/tit/BerlekampMT78,DBLP:conf/isit/MisoczkiTSB13}.
Given a $(v,w)$-regular binary code with a $r$$\times$$n$ parity-check matrix
$H$ and the value of a syndrome $s$$=$$He^{\mathtt{T}}$ derived from an
unknown error vector $e$ with weight $t$$=$$\weight{e}$, the corresponding set of
parity-check equations can be written as $ \sum_{j=0}^{n-1} h_{i,j} e_j = s_i$,
where $e_j$ are the unknown error bits, $h_{i,j}$ are the known coefficients of the $H$
matrix, $s_i$ the known constant terms provided as bits of the syndrome, while additions
and multiplications are meant to be executed modulo $2$ (i.e., as operations in the Galois
Field GF($2$)).
Any equation is said to be either {\em satisfied}, if its constant
term is clear (i.e., $s_i$$=$$0$), or {\em unsatisfied}, if its constant
term is asserted (i.e., $s_i$$=$$1$).
The decoding strategy of interest in this work is the (parallel) bit flipping
decoding procedure introduced by Gallager in~\cite{DBLP:journals/tit/Gallager62}
to iteratively estimate the most likely value $\bar{e}$ of the error vector
$e$, given $s$ and $H$ and the initial value $\bar{e}=0_{1\times n}$ as shown
in Algorithm~\ref{algo:bitflipping}.
\begin{algorithm}[!t]
  \DontPrintSemicolon
  {\scriptsize
  \KwIn{%
  $s$: $r\times 1$ syndrome (obtained as $s=H\widetilde{c}^{\mathtt{T}}=He^{\mathtt{T}}$), where\newline
  $\phantom{s: r \times 1} \widetilde{c}$ is an error affected codeword,
  and $e$ is the $1 \times n$ unknown\newline
  $\phantom{s: r \times 1}$error vector with weigth $t$,\newline
  $H$: $r\times n$ parity-check matrix,\newline
  $\mathtt{iterMax}$: max number of of permitted iterations.
  }
  \KwOut{$\bar{e} = [\bar{e}_0, \ldots, \bar{e}_{n-1}]$: bits of the error
  vector estimate,\newline
  $\mathtt{decodeOk}$: value indicating success, $1$, or fail, $0$.
  }
  $\bar{e} \gets \mathtt{0}_{1 \times (n-1)}$ \tcp*[l]{\scriptsize bit vector }
  $\mathtt{iter} \gets 1$\;
  \While{($s \neq 0_{r\times 1}$ \textbf{\em and}
          $\mathtt{iter} \leq \mathtt{iterMax}$)}{
        \For{$j$ \textbf{\em from} $0$ \KwTo $n-1$}{
           $\upc \gets \langle s, h_{:,j}\rangle$\tcp*[l]{
           \scriptsize $s_i$$\cdot$$h_{i,j}$ as integer}
        }
     $\mathtt{th} \gets \text{\sc ThresholdChoice}(\mathtt{iter},s)$\;
     \For{$j$ \textbf{\em from} $0$ \KwTo $n-1$}{
        \If{($\mathtt{upc}_j \geq \mathtt{th}$)}{
                 $\bar{e}_j \gets \bar{e}_j\oplus 1$\;
                 $s \gets s \oplus h_{:, j}$\;
        }
     }
     $\mathtt{iter} \gets \mathtt{iter} + 1$\;
  }
  \lIf{($s = 0_{r\times 1}$)}{$\mathtt{decodeOk} \gets 1$}
  \lElse{$\mathtt{decodeOk} \gets 0$}
  \KwRet{$\bar{e}, \mathtt{decodeOk}$}\;
  }
  \caption{{\sc Bit Flipping Algorithm}\label{algo:bitflipping}}
\end{algorithm}
It proceeds by iteratively refining an estimate
$\bar{e}$ of the value of the error vector.
In particular, denoting with $\bar{d}=e\oplus\bar{e}$,
the vector of \emph{discrepancies} between the estimate and the actual
error vector, we can observe that Algorithm~\ref{algo:bitflipping} keeps
the relation $H\bar{d}^{\mathtt{T}}=s$ true at the end of each iteration.
In the following, we will refer to a value at the end of any iteration
of the loop at lines $3$--$11$ adding a superscript integer with the
value of the variable $\mathtt{iter}$ between round braces,
e.g., $s^{(1)}$ will denote the value of the syndrome vector at the end
of the first iteration.
A null superscript will denote values before the execution of
Algorithm~\ref{algo:bitflipping}, e.g., $s^{(0)}$ will denote the initial
syndrome value taken as input.

This work builds on the one from~\cite{DBLP:conf/isit/AnnechiniBP24}, where
the authors denotes as $\mathcal{E}^{(\mathtt{iter})}$ the random
variable modeling the Hamming weight $\discre$$\in$$\{0,\ldots, n\}$
of the discrepancy vector $\bar{d}^{(\mathtt{iter})}$ after the
$\mathtt{iter}$-th iteration of the decoding algorithm.
They provide a closed form method to compute its probability
mass function (p.m.f.), $\Pr(\mathcal{E}^{(\mathtt{iter})}$$=$$\discre)$,
for $\mathtt{iter}$$=$$1$ and $\mathtt{iter}$$=$$2$,
starting from $\Pr(\mathcal{E}^{(\mathtt{0})}$$=$$t)$$=$$1$
(since $\bar{e}^{(0)}$ is null and $\weight{e} = t$).

In particular, denoting as $\mathcal{W}_t$ the random variable modeling
the initial weight of the syndrome, i.e., $y =\weight{s^{(0)}}$,
the contribution in~\cite{DBLP:conf/isit/AnnechiniBP24} shows how to
compute the p.m.f. $\Pr(\mathcal{W}_t = y)$ as a function of the code
parameters and weight $t$ of the error vector, and evaluate
$\Pr(\mathcal{E}^{(\mathtt{iter})}=\discre)$ as:\\[-2.7em]
\begin{center}
\resizebox{1.0\linewidth}{!}{
$
\hspace{-0.5em}\Pr(\mathcal{E}^{(\mathtt{iter})}=\discre)$$=$
$\displaystyle \sum_{y = 0}^{r}
\condprob{\mathcal{E}^{(\mathtt{iter})} = \discre}{\mathcal{W}_t = y}
\Pr(\mathcal{W}_t = y).
$
}
\end{center}
The computation in~\cite{DBLP:conf/isit/AnnechiniBP24} considers
the {\sc ThresholdChoice} function as yielding values chosen a-priori,
only depending on $\mathtt{iter}$, with no dependency on the syndrome value
$s$ or other runtime values of the decoder.
It is however known to be beneficial to the decoding capability to take
into account information such as the syndrome weight into the choice
of a threshold~\cite{DBLP:phd/hal/Chaulet17,LEDA}.
The rationale of such an approach is that the syndrome weight provides
information on how many discrepancies are left over to correct, and on the
extent of the interference on the parity-check equations happening due to the
positions of the erroneous bits in $e$.
Picking the thresholds as a function of the weight of one or more syndromes
(in distinct iterations of the decoder) has
proven to be of particular interest and efficiency in LDPC/MDPC iterative
decoding for cryptographic purposes.
In particular, the post-quantum BIKE~\cite{BIKE} and LEDAcrypt
cryptosystems~\cite{LEDA} employed such an approach.
The former, in its last revision, did so to improve on the
DFR of the system~\cite{DBLP:phd/hal/Chaulet17}, which was found not to be
meeting the requirements to guarantee proper security levels~\cite{WangWW23}.
The latter aimed to reduce the DFR of the decoder when the cryptosystem
is used for ephemeral key encapsulation.
\newline\noindent\textbf{Contributions.} Providing a closed form model for
the overall DFR of iterative decoders, which apply a threshold choice function
that is dependent on the whole state of the algorithm, is the open research
challenge this work addresses, having both a theoretical and a practical 
interest in cryptosystem 
design~\cite{LEDA,DBLP:conf/cbc/BaldiBCPS21,DBLP:conf/icete/BaldiBCPS20a,BIKE}.
To do so, we introduce a model for the distribution of the weight of the
syndrome after the first iteration of the bit flipping decoder,
$\weight{s^{(1)}}$.
Given such a distribution, we derive the threshold values for the first and
second iteration minimizing the expected DFR value, which is computed as
a function of the code parameters, the error weight, the syndrome weights
and the threshold values themselves.
While the proposed approach allows to perform a joint optimization of
the threshold values for the first and second iteration, it also allows
to perform such threshold value optimizations separately, in turn
allowing our proposal to scale up to code sizes of interest for
cryptographic purposes, while operating the computation on an off-the-shelf
laptop.
We provide numerical validations for our model of the distribution of
the syndrome weight $\weight{s^{(1)}}$ and for the fitness of our
two-iterations DFR model derived from it.
We quantify the two-iteration DFR performance gap of our threshold selection
criteria w.r.t the current state of the art one employed in the
BIKE cryptosystem, and a fixed threshold decoder.
Furthermore, we show that our approach matches the performance of the
one obtained by regressing the best possible thresholds from
decoding of $10^8$ randomly chosen errors with fixed weight.
For the sake of reproducibility, our codebase from which we obtained
our results is available at~\footnote{\url{https://crypto.deib.polimi.it/threshold_optimization.zip}}
%

%% file: background.tex
In the following, before deriving the distribution of the weight of the
syndrome computed at the end of the first iteration of the decoder
(Section~\ref{sec:syn2it}), showing how to compute in closed form
the DFR of the decoder at the end of its second iteration
and derive the thresholds required for the
evaluation of the unsatisfied parity-check counters ($\mathtt{upc}$) in
Algorithm~\ref{algo:bitflipping} (Section~\ref{sec:model}),
we summarize the main concepts and notation of the analyses developed
in~\cite{DBLP:conf/isit/AnnechiniBP24} upon which we build.
In particular, we assume the same premises, that is we consider the rows of
$H$ as independently and uniformly random drawn from the set of binary vectors
having length $n$ and weight $w$, keeping into account the constraint on the
weight of each column to be $v$.
Furthermore, we adopt the same logical partition of the bit positions in
the discrepancy vector after the first iteration (i.e., $\bar{d}^{(1)}$)
that was introduced in~\cite{DBLP:conf/isit/AnnechiniBP24}, and each bit
position, $i$, of $\bar{d}^{(1)}$ is thought to be included
in a set labeled with a pair $(a,b)$, where $a=e_i$ and $b=\discvectoridx{i}$.
The four resulting classes and their cardinalities are denoted as
$\jset{a}{b}$ and $|\jset{a}{b}| = \epsilon_{ab}$, respectively.
Differently from~\cite{DBLP:conf/isit/AnnechiniBP24}, where the thresholds
employed in Algorithm~\ref{algo:bitflipping} are chosen a-priori, we derive
the criterion to compute the threshold applied during the
first iteration of Algorithm~\ref{algo:bitflipping}, as a
function of the weight of the received syndrome $y=\weight{s^{(0)}}$,
i.e., $\thfirstit$, while
the threshold to be applied during the second iteration is derived as a
function of $y$, the amount $z_0$ of satisfied
parity-check equations that \underline{became unsatisfied} after the first
iteration, and the amount $z_1$ of equations that \underline{remained unsatisfied}
after the first iteration, i.e., $\thsecondit$.
Note that $z_0+z_1=\weight{s^{(1)}}$, and that computing $z_0$ and $z_1$
requires little computational effort.
Our model allows also to derive the {\sc ThresholdChoice} procedure as a
function of other informative quantities about the state of the decoder,
such as the number of flips that take place in the estimated error vector
at the end of the previous iteration, or the values of each unsatisfied
parity-check counter, $\mathtt{upc}_j, 0$$\leq$$j$$\leq$$n$$-$$1$ at the end
of the previous iteration (thus, including the behavior of decoders such as
the one in~\cite{DBLP:conf/pqcrypto/DruckerGK20}).
However, in our study of the two-iterations bit flipping decoder, we chose
to employ $z_0$ and $z_1$ only, for the sake of computational efficiency and the
relatively small memory footprint needed to tabulate the values
yielded by the {\sc ThresholdChoice} function.

For the analyses developed in the next sections, we are going to re-use the
the derivation in~\cite{DBLP:conf/isit/AnnechiniBP24} that leads to the
computation of the probabilities $\punsatz$, and $\punsato$, which in turn
capture the likelihood of any parity-check equation in $He^{\mathtt{T}} = s$,
(e.g., the $i$-th one, $0$$\leq$$i$$\leq$$r$$-$$1$), to be unsatisfied, given
that either a clear bit or an asserted bit, in the unknown
actual error vector (say it $e_j =0$, or $e_j = 1$) appears among the terms of
the equation itself (i.e., $h_{i,j} =1$).
In this work, we are going to consider $\punsatz$ and $\punsato$ as a function 
of the weight of the original syndrome $y$$=$$\weight{s^{(0)}}$.
Subsequently, these quantities are employed to compute the statistical
distribution of the $\mathtt{upc}$ values, as a function of $\thfirstit$,
at the end of the first iteration of the decoder, and derive from them the
probabilities $p_{\mathtt{flip}|0}$ and $p_{\mathtt{flip}|1}$ that a bitflip in
the error vector estimate $\bar{e}$ have been incorrectly ($\discplus$) or
correctly ($\discminus$) performed, i.e., if it has increased or decreased
the weight of the discrepancy vector, respectively:
$$
\begin{array}{l}
p_{\mathtt{flip}|0} = \sum_{a=\thfirstit}^v \bindist{v}{\punsatz}{a},\\
p_{\mathtt{flip}|1} = \sum_{a=\thfirstit}^v \bindist{v}{\punsato}{a}.
\end{array}
$$
The previous formulae, where $\bindist{\mathtt{tr}}{\mathtt{spr}}{\mathtt{ns}}$
denotes the binomial p.m.f., with $\mathtt{ns}$ success events out of
$\mathtt{tr}$ independent events and a success probability of $\mathtt{spr}$,
are consequently employed to compute the p.m.f. of the number of discrepancies
added or substracted to the weight of $\bar{d}^{(1)}$, as follows:
$$
\begin{array}{l}
\deltapy{\discplus} = \bindist{n-t}{p_{\mathtt{flip}|0}}{\discplus}\\
\deltamy{\discminus} = \bindist{t}{p_{\mathtt{flip}|1}}{\discminus},
\end{array}
$$
pointing out the dependency from the original syndrome weight $y$.
Further consequences are that, at the end of the first decoding iteration,
$\prob{|\jset{0}{1}| = m}$ and $\prob{|\jset{1}{1}| = m}$,
can be expressed as:
\begin{align*}
\epsilon_{01} = |\mathbf{J}_{0,1}|,\ & \prob{\epsilon_{01} = m} = \deltapy{m},\\
\epsilon_{11} = |\mathbf{J}_{1,1}|,\ & \prob{\epsilon_{11} = m} = \deltamy{t-m}.
\end{align*}
Finally, the probabilities of flipping ($p_{\mathtt{flip}|\ldots}$) or
maintaining ($p_{\neg \mathtt{flip}|\ldots}$) a bit in the error
vector estimate given that the corresponding bit in the actual
error vector (clear, i.e., $p_{\neg \mathtt{flip}|0\ldots}$ or asserted,
i.e., $p_{\neg \mathtt{flip}|1,\ldots}$) is involved in at least one out of $v$
parity-check equations, either satisfied or unsatisfied are:
$$
\begin{array}{l}
\pflipz{\mathtt{oneEqSat}} = \sum_{a=\thfirstit}^{v-1} \bindist{v-1}{\punsatz}{a}\\
\pnoflipo{\mathtt{oneEqSat}} = \sum_{a=0}^{\thfirstit-1} \bindist{v-1}{\punsato}{a}\\
\pflipz{\mathtt{oneEqUnsat}} = \sum_{a=\thfirstit-1}^{v-1} \bindist{v-1}{\punsatz}{a}\\
\pnoflipo{\mathtt{oneEqUnsat}} = \sum_{a=0}^{\thfirstit-2} \bindist{v-1}{\punsato}{a}.
\end{array}
$$

%% file: syndrome_weight_distribution.tex
In the following, we describe how to compute the distribution of the random
variable $\synwsit{\epsilon_{01}}{\epsilon_{11}}$ that models the weight
of the syndrome vector at the end of the first iteration of the decoding
algorithm, which in turn performed $\discplus$$=$$\epsilon_{01}$ incorrect
flips and $\discminus$$=$$t$$-$$\epsilon_{11}$ correct flips of the bit
values into the estimated error vector $\bar{e}$.
In particular, the p.m.f $\Pr(\synwsit{\epsilon_{01}}{\epsilon_{11}} = z)$,
$z$$\in$$\{0,\ldots,r\}$, is derived as a function of the code parameters
$n, k, v, w$ and the weight $t$ of the error vector.
For the sake of clarity, in this section we are going to omit the conditioning
of every event to the one stating that the random variable modeling the
original syndrome weight takes a specific value, i.e., the event
($\mathcal{W}^{(0)}_t = y$), where $y = \weight{s^{(0)}}$.
We denote with $(\mathcal{Z}_0,\mathcal{Z}_1)$ the pair of random variables
with joint p.m.f. $\Pr(\mathcal{Z}_0 = z_0, \mathcal{Z}_1 = z_1)$, where
$0$$\leq$$z_0$$<$$r$, $0$$\leq$$z_1$$\leq$$r$, and $0$$\leq$$z_0$$+$$z_1$$\leq$$r$.
These variables model the number of parity-check equations which became
(resp. stayed) unsatisfied after computing the first iteration.
We note that $s^{(1)}$ is computed as the sum of the columns of $H$ indexed 
by asserted bits in $\discvector$.
Assuming to know $\epsilon_{01}$ and $\epsilon_{11}$, we define a
time-dependent non homogeneous Markov process, which models the procedure
computing $s^{(1)}$ yielding the same result of the operations in
lines $7$--$10$ in Algorithm~\ref{algo:bitflipping}.
The positions of the (correct and incorrect) asserted bits in
$\discvector$ belong to $\jset{0}{1}$ and $\jset{1}{1} $, and we
consider different transition probabilities depending on whether the position
in $\bar{d}^{(1)}$ inducing the column addition being modeled is in one set
or the other.
For all possible values of $l_{01}$$\in$$\{0,\ldots,\epsilon_{01}\}$,
$l_{11}$$\in$$\{0,\ldots,\epsilon_{11}\}$, we consider the states
of the Markov process as the elements of the sequences
$(\discvector,s^{(1)})_{0,0}$, $\ldots$,
$(\discvector,s^{(1)})_{l_{01},l_{11}}$, $\ldots$,
$(\discvector,s^{(1)})_{\epsilon_{01},\epsilon_{11}}$
to represent all possible chains of additions of columns
of $H$ to $s^{(1)}$
that may be envisioned in correspondence with the bit values
in $\discvector$. The single initial state of the said sequences
is defined by
a null discrepancy vector and a null syndrome vector.
The initial distribution of our Markov process is:
$$
\Pr\left(\left(\mathcal{Z}_0 = z_0, \mathcal{Z}_1 = z_1\right)_{(0,0)}\right) =
\begin{cases}
    1 & \text{if } z_0=z_1=0\\
    0 & \text{elsewhere}
\end{cases}
$$
Given the commutativity of additions, all the
paths to reach the final state can be considered equivalent.
As a consequence, to ease the computations and without loss of generality
we assume to perform all the steps
$(0,0) \to (1,0) \to \cdots \to (l_{01},0)\to \cdots \to (\epsilon_{01},0)$,
incrementing $l_{01}$,
then performing all the steps in $(\epsilon_{01},0) \to
(\epsilon_{01},1) \to
\cdots \to (\epsilon_{01},l_{11})\to \cdots \to (\epsilon_{01},
\epsilon_{11})$, incrementing $l_{11}$.
Our aim in defining this Markov process is to compute the final joint distribution
of $(\mathcal{Z}_0,\mathcal{Z}_1)_{(\epsilon_{01},\epsilon_{11})}$.
To this end, we derive the $(r+1)r$$\times$$(r+1)r$ transition
matrix of the process containing in each cell the values:\\[0.5em]
\noindent\resizebox{1.0\linewidth}{!}{$
\Pr\left(\left(\mathcal{Z}_0 = x_0, \mathcal{Z}_1 =
x_1\right)_{(l_{01}+1,l_{11})} | \left(\mathcal{Z}_0 =
z_0, \mathcal{Z}_1 = z_1\right)_{(l_{01},l_{11})} \right),$}\\[0.5em]
\noindent\resizebox{1.0\linewidth}{!}{$\Pr\left(\left(\mathcal{Z}_0 = x_0, \mathcal{Z}_1 = x_1
\right)_{(l_{01},l_{11}+1)} | \left(\mathcal{Z}_0 = z_0,
\mathcal{Z}_1 = z_1\right)_{(l_{01},l_{11})} \right),$}\\[0.5em]
for all $l_{01} \in \{0,\ldots,\epsilon_{01}\}$, $l_{11}\in
\{0,\ldots,\epsilon_{11}\}$, and all $x_0,x_1$ in the same range
of $\mathcal{Z}_0,\mathcal{Z}_1$, respectively.
The complete derivation is
reported in Appendix~\ref{app:trmatrix} of the extended version of
this paper~\cite{arxivextended}.
Representing the joint bivariate p.m.f. distribution, at each step of
the process, as a $(r+1)r$ array, the application of the transition matrix
along the paths described before allows to exhibit the distribution on
the final state, i.e.,
$\Pr(\mathcal{Z}_0 = z_0,\mathcal{Z}_1 = z_1)_{(\epsilon_{01},\epsilon_{11})}$,
for all admissible values of $z_0$ and $z_1$.
At the end of the first iteration fo the decoder, knowing that
$|\jset{0}{1}|$$=$$\epsilon_{01}$ and
$|\jset{1}{1}|$$=$$\epsilon_{11}$, it is possible to compute
the p.m.f. $\Pr(\synwsit{\epsilon_{01}}{\epsilon_{11}} = z)$
composing the following with the formula of total probabilities
to keep into account the conditioning by the event
($\mathcal{W}^{(0)}_t = y$).
$$
\sum_{z_0 = \max(0,z-y)}^{\min(z,r-y)}
\Pr\left(\left(\mathcal{Z}_0 = z_0, \mathcal{Z}_1 = z - z_0\right)_{
(\epsilon_{01},\epsilon_{11})}\right).
$$

%% file: dfrmodel.tex
\begin{figure*}[!t]
\begin{center}
    \input{figures/tikz_pic_syn2it}
\end{center}
\vspace{-0.5cm}
    \caption{Numerical validation of 
 the model of syndrome weight distribution after the first iteration.
 Code parameters: $n=9,602$, $\frac{k}{n}=\frac{1}{2}$, $v=45$,
 $t=50$, $\mathtt{th}^{(1)}=25$ fixed.
 Numerical results obtained with $10^8$ random samples.
 \ref{mark:sinweven} depicts the model distribution
 over even values of $z$, \ref{mark:sinwodd} over odd ones.}
    \label{fig:syn2it}
\end{figure*}
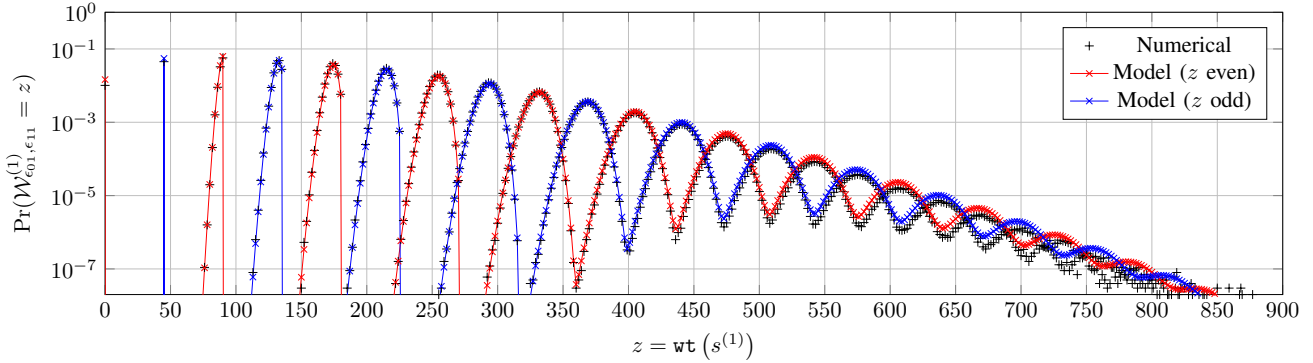

To derive the DFR after the second iteration
of the parallel decoder, we derive the probabilities of flipping bits
of $\discvector$ depending on the inclusion of their positions
in $\jset{a}{b}$, with $a, b$$\in$$\{0,1\}$, as a function of
$y=\weight{s^{(0)}}, \epsilon_{01}, \epsilon_{11}, z_0, z_1$, which
summarize the state of the decoder after its first iteration.
The said DFR  equals the probability that the decoder makes
\textit{at least} one incorrect decision during its second iteration.
In particular, we derive the $\pflipzz$, $\pflipzo$, $\pflipoz$ and
$\pflipoo$, as the probability of flipping bits in
$\jset{0}{0}$, $\jset{0}{1}$, $\jset{1}{0}$, and $\jset{1}{1}$,
respectively, as
reported in Appendix~\ref{app:flipprob} of the extended version of this
paper~\cite{arxivextended}.
Consequently, we can derive
$\mathtt{DFR}(y,\epsilon_{01},\epsilon_{11},z_0,z_1)$ as the probability of
\textit{not} performing all the right flips during the second iteration,
i.e.:
$$
1 - (1 - \pflipzz)^{n-t-\epsilon_{01}}
\cdot
\pflipzo^{\epsilon_{01}}
\cdot
(1-\pflipoz)^{t-\epsilon_{11}}
\cdot
\pflipoo^{\epsilon_{11}}
$$
To obtain a closed form of the DFR
as a function of only code parameters and the original error
weigth $t$, denoted as, $\mathtt{DFR}^{(2)}$, we proceed by
enriching our computational procedure with the averaging over
all possible values of $y, \epsilon_{01}, \epsilon_{11}, z_0, z_1$.
$$
\begin{array}{l}
\mathtt{DFR}^{(2)} =        \sum_{y=0}^{r} \Pr\left( \mathcal{W}^{(0)}_{(t)} = y\right) \cdot\\
         \cdot \left( \sum_{\epsilon_{01}=0}^{n-t} \sum_{\epsilon_{11}=0}^t
\deltapy{\epsilon_{01}} \cdot \deltamy{t-\epsilon_{11}} \cdot \right.\\
         \cdot \left( \sum_{z_0 = 0}^{r-y} \sum_{z_1 = 0}^{y}
\Pr\left(\left(\mathcal{Z}_0 = z_0, \mathcal{Z}_1 = x - z_0\right)_{(\epsilon_{01},\epsilon_{11})}\right) \right. \cdot\\
\left. \left. \cdot \mathtt{DFR}(y,\epsilon_{01},\epsilon_{11},z_0,z_1)\right)\right)
\end{array}
$$
The aforementioned step, in line with the assumption stated in 
Section~\ref{sec:background}, and in~\cite{DBLP:conf/isit/AnnechiniBP24}, 
considers each flipping decision in isolation.
We provide an extended discussion on taking into account the whole
set of constraints coming from the regularity of the parity
check matrix $H$ at once, in Appendix~\ref{app:regulardfr}
of the extended version of this paper~\cite{arxivextended}.

\noindent\textbf{Computation of Thresholds.}
We have now reached the point where we want to find the most convenient
threshold selection functions $\optthfirstit$ and $\optthsecondit$. To this
end, we construct them pointwise as $\mathtt{th}^{(1)}(a),
\mathtt{th}^{(2)}(a,b,c)$ for all possible fixed values $a,b,c$. 
To express our procedure, we highlight the
dependency of $\mathtt{DFR}$ on the threshold choice as
$\mathtt{DFR}(a,\epsilon_{01},\epsilon_{11},b,c,\mathtt{th}^{(1)}(a),
\mathtt{th}^{(2)}(a,b,c))$, and likewise the one of the discrepancy
distributions after the first iteration as
$\delta_{+,a}(\epsilon_{01},\mathtt{th}^{(1)}(a)),
\delta_{-,a}(t-\epsilon_{11},\mathtt{th}^{(1)}(a))$.
We determine each pair of values $\big(\widehat{\mathtt{th}}^{(1)}(a),
\widehat{\mathtt{th}}^{(2)}(a,b,c)\big)$
as:
\resizebox{\linewidth}{!}{
$
\displaystyle
\argmin_{
{\scriptsize \begin{array}{l}
0\leq \tau_1=\mathtt{th}^{(1)}(a) \leq v; \\
0\leq \tau_2=\mathtt{th}^{(2)}(a,b,c) \leq v
\end{array}
}} 
\Big\{\sum_{\epsilon_{01}=0}^{n-t} \sum_{\epsilon_{11}=0}^t
\delta_{+,a}(\epsilon_{01},\tau_1)\, \delta_{-,a}(t-\epsilon_{11},\tau_1) \cdot
$
}
\resizebox{\linewidth}{!}{
$
\cdot\Pr\Big(\left(\mathcal{Z}_0 = b, \mathcal{Z}_1 = c\right)_{(\epsilon_{01},\epsilon_{11})} \Big) \cdot \mathtt{DFR}(a,\epsilon_{01},\epsilon_{11},b,c,\tau_1,\tau_2)
\Big\}. $
}

Deriving all the points of the $\optthfirstit$ and $\optthsecondit$
functions proves to be computationally expensive, especially for cryptographic-grade code parameters, as it scales as $r^3nt$.
To reduce the computational effort, we split the joint optimization problem
into two separate ones.
We first find the $r$ values constituting $\optthfirstit$ employing only
information available at the end of the first iteration.
In particular, for each value $0$$\leq$$a$$\leq$$r$, we compute the
threshold value, which minimizes the amount of discrepancies left by the
first iteration,
i.e., $\widehat{\mathtt{th}}^{(1)}(a)$ is:
\resizebox{\linewidth}{!}{
$\displaystyle \argmin_{
{\scriptsize \begin{array}{l}
0\leq \tau_1=\mathtt{th}^{(1)}(a) \leq v \\
\end{array}
}} \left\{
\sum_{\epsilon_{01}=0}^{n-t}
\epsilon_{01}
 \delta_{+,a}(\epsilon_{01},\tau_1)
+
\sum_{\epsilon_{11}=0}^t
\epsilon_{01}
 \delta_{-,a}(t-\epsilon_{01},\tau_1)
\right\}
$.
}

Once $\optthfirstit$ is available, we optimize the second iteration threshold as a function of $z_0$ and $z_1$ alone, i.e. $\optthseconditsimple$. To find the values $\widehat{\mathtt{th}}^{(2)}(b,c)$, for all the pairs
of values $b,c$
we minimize the DFR after the second iteration
while employing each value obtained with the first iteration threshold
selection function $\widehat{\mathtt{th}}^{(1)}(y)$, averaged over 
every possible value of $y$:

\vspace{0.2cm}
$
\begin{array}{l}
{\displaystyle
\widehat{\mathtt{th}}^{(2)}(b,c) = 
\argmin_{
{\scriptsize 
0\leq \tau_2=\mathtt{th}^{(2)}(b,c) \leq v
}}}
\Big\{
\sum_{y=0}^r
\Pr\left( \mathcal{W}^{(0)}_{(t)} = y\right) \cdot
\end{array}
$
\resizebox{\linewidth}{!}{
$
\begin{array}{l}
\cdot \left(
\sum_{\epsilon_{01}=0}^{n-t} \sum_{\epsilon_{11}=0}^t
\delta_{+,y}(\epsilon_{01},\widehat{\mathtt{th}}^{(1)}(y))\, \delta_{-,y}(t-\epsilon_{11},\widehat{\mathtt{th}}^{(1)}(y)) \cdot \right.
\end{array}
$
}
\resizebox{\linewidth}{!}{
$
\begin{array}{l}
\left. \cdot\Pr\Big(\left(\mathcal{Z}_0 = b, \mathcal{Z}_1 = c\right)_{(\epsilon_{01},\epsilon_{11})} \Big) \cdot \mathtt{DFR}(y,\epsilon_{01},\epsilon_{11},b,c,\widehat{\mathtt{th}}^{(1)}(y),\tau_2)
\right) \Big\}
\end{array}
$
}
With this strategy, we are able to compute $\optthfirstit$ and $\optthseconditsimple$ for cryptographic-grade code parameters, while requiring a small memory footprint for tabulating the thresholds to be used by the {\sc ThresholdChoice} function.


%% file: figures/tikz_pic_syn2it.tex
\begin{tikzpicture}[scale=0.85]
  \begin{axis}[
                width=20cm, 
                height=6cm,
               xmin = 0,
               xmax = 900,
               grid = major,
               ymode= log,
               ymax = 1,
               ymin = 2e-8,
               ytick={1,1e-1,1e-3,1e-5,1e-7},
               legend style={at={(0.9,0.62)},anchor=south},
               mark size=2pt,
               xlabel={$z = \weight{s^{(1)}}$},
               ylabel={$\Pr(\mathcal{W}^{(1)}_{\epsilon_{01},\epsilon_{11}}=z)$},
               line width=0.2pt,
               mark options=solid]

    \addplot[black,only marks, mark=+]
             table [x = w,
                    y expr = \thisrow{sim}, 
                    col sep = comma]{data/syn_w_2it_n0_2_r_4801_v_45_t_50_th_25_10e8.csv};
    \addlegendentry{Numerical};
    \addplot[red, mark=x]
             table [x = w,
                    y expr = \thisrow{model}, 
                    col sep = comma]{data/syn_w_2it_even_n0_2_r_4801_v_45_t_50_th_25_10e8.csv}; \label{mark:sinweven}
    \addlegendentry{Model ($z$ even)};
    \addplot[blue, mark=x]
             table [x = w,
                    y expr = \thisrow{model}, 
                    col sep = comma]{data/syn_w_2it_odd_n0_2_r_4801_v_45_t_50_th_25_10e8.csv};\label{mark:sinwodd}
    \addlegendentry{Model ($z$ odd)};

\end{axis}
\end{tikzpicture}

%% file: expres.tex
\begin{figure*}[!t]
    \hspace{-1.2em}
    \subfloat[\label{fig:density_sweep}]{
       \input{figures/tikz_pic_density_sweep}
    }
    \hspace{-0.3em}    
    \subfloat[\label{fig:density_sweep_comparison}]{
       \input{figures/tikz_pic_density_comparison}
    }
    \hspace{-0.3em}    
    \subfloat[\label{fig:exhaustive_comparison}]{
       \input{figures/tikz_pic_exhaustive_comparison}
    }
\caption{$\mathtt{DFR}^{(2)}$ for $(v,2v)$-regular LDPC codes
  with $\frac{k}{n}=\frac{1}{2}$, parallel decoder. Each data point is obtained
  with either $10^8$ decodes or reaching $100$ decoding failures.
  (a) Using our thresholds, our $\mathtt{DFR}^{(2)}$ model (solid lines) vs.
  numerical simulations (crosses); (b) Modeled $\mathtt{DFR}^{(2)}$: our thresholds
  (solid lines), majority thresholds
  $\mathtt{th}^{(1)}=\mathtt{th}^{(2)}=\lceil \frac{v+1}{2}\rceil$ (dashed lines);
  (c) for $v=19$ and $t=18$, performance comparison among majority
  thresholds (blue crosses), our choice (red crosses), best choice obtained evaluating a
  posteriori $10^8$ decodes on different syndromes (black circles)}
\end{figure*}
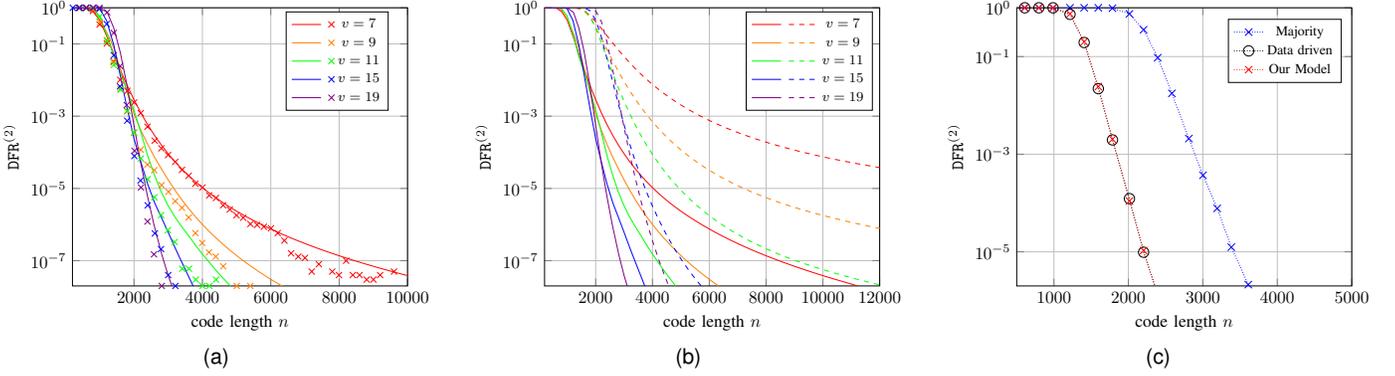

In this section, we provide numerical validations of our threshold
choice approach and second iteration DFR model.
The first observation we make is that the values of $\optthfirstit$
grows essentially linearly in the syndrome weight,
and our $\optthfirstit$ matches the one which can be numerically regressed
from simulations.
An in depth discussion and plots are reported in
Appendix~\ref{app:additional_results} of~\cite{arxivextended}.

Figure~\ref{fig:syn2it} depicts the numerical validation of our model for 
the distribution of the syndrome weight after one iteration
$\Pr(\mathcal{W}^{(1)}_{\epsilon_{01},\epsilon_{11}}=z)$, averaged
over all possible values of $\epsilon_{01}$$\in$$\{0,\ldots,n-t\},\epsilon_{11}$$\in$$
\{0,\ldots,t\}$, on a
cryptographic grade parameter set from LEDAcrypt~\cite{LEDA}.
The shape of the distribution and the goodness of
fit of our model is best understood observing that the weight of a 
syndrome $s^{(0)}$ (or at any other iteration), computed on an discrepancy vector
of weight $t$, has the same \textit{parity} as $t \cdot v$.
Indeed, the syndrome is computed adding
$t$ columns, and each one causes $v$ bit flips onto the syndrome. Each
bit flip changes the parity of the syndrome vector, therefore the parity
of the syndrome (starting at zero), is changed $t\cdot v$ times. 
After the first iteration, the parity of 
$\mathcal{W}^{(1)}_{\epsilon_{01},\epsilon_{11}}$ matches the one of
$(\epsilon_{01}+\epsilon_{11}) \cdot v$. Thus, 
when $v$ is odd, $\Pr(\mathcal{W}^{(1)}_{\epsilon_{01},\epsilon_{11}}=z)$ can be
logically split between even and odd values (resp. red and blue 
in Figure~\ref{fig:syn2it}): our model provides a very good fit of both cases. 

Figure~\ref{fig:density_sweep} compares numerical DFR values of a
two-iterations parallel bit-flipping decoder with our estimation technique, 
while varying either the code density and sweeping over the code length range 
$n$$\in$$\{200,...,12000\}$. The thresholds employed for the two iterations
are chosen with respect to our strategy from Section~\ref{sec:model}.
As it can be seen, our two-iteration DFR estimation technique provides a 
reliable estimate of the waterfall region, and a conservative estimate for 
the floor.

Figure~\ref{fig:density_sweep_comparison} depicts, for the same parameter
sets as Figure~\ref{fig:density_sweep}, a comparison between the estimated
DFR of the parallel decoder employing our $\optthfirstit,\optthsecondit$ (solid lines)
against the DFR of a decoder employing fixed majority thresholds (dashed lines).
As it can be seen, our threshold choice improves the DFR by about $10^3\times$
across the board.

Figure~\ref{fig:exhaustive_comparison} reports a comparison on numerical simulations
of a two iteration decoder with our $\optthfirstit,\optthsecondit$ (red)
with respect to a majority threshold decoder (blue), and a decoder (black)
where we obtained the thresholds by performing $10^8$ decoding actions with
all possible thresholds and selected the $\optthfirstit,\optthsecondit$ that empirically
minimized the $\mathtt{DFR}^{(2)}$. The decoder was then tested on $10^8$ freshly
randomly selected inputs. As it can be seen, employing our threshold selection
criterion matches the decoding performance of the best possible one
obtained regressing it from the data.
\begin{figure}[!t]
\begin{center}
    \input{figures/tikz_pic_bike_avg_err} 
\end{center}
\vspace{-0.5cm}
    \caption{
    Average number of discrepancies left after two iteration for QC-MDPC codes
  with rate $\frac{k}{n}=\frac{1}{2}$, $v=71$, $t=134$, parallel decoder 
  employing three different threshold policies: majority voting with margin 
  $\delta=3$ ($\mathtt{th}^{(1)}=\mathtt{th}^{(2)} =\lceil \frac{v+1}{2} 
  \rceil + \delta$), BIKE-flip threshold selection~\cite{BIKE}, and
  thresholds computed by our model. Each data point obtained performing either 
  $10^6$ decoding actions, or reaching $10^6$ total discrepancies. $r=12,323$
  corresponds to the parameter set of
  BIKE for NIST security level 1.\label{fig:bike_comparison}}
\end{figure}
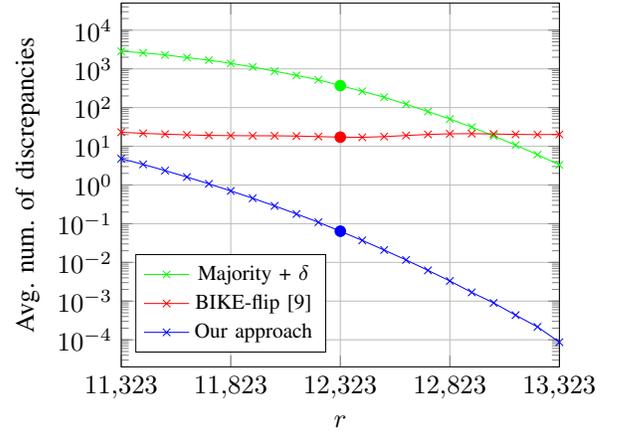
Figure~\ref{fig:bike_comparison} shows a comparison between the
threshold selection procedure employed in BIKE~\cite{BIKE} (red), comparing it 
both to our approach (blue) and a majority decodeer with margin $\delta=3$
(i.e., with the same margin which BIKE adds to its own syndrome-weight dependent
threshold selection function).
The figure shows that our method effectively reduces the expected number of 
leftover discrepancies after two iterations, $\weight{\bar{d}^{(2)}}$, by
a factor of $\approx 3\cdot 10^{2}$ w.r.t.~\cite{BIKE} and $> 5\cdot 10^{3}$
w.r.t. the majority approach. Results for the BIKE parameters for other security
levels are available in Appendix~\ref{app:additional_results}
of~\cite{arxivextended}.

%% file: figures/tikz_pic_density_sweep.tex
\pgfkeys{/pgf/number format/.cd,1000 sep={}}
\begin{tikzpicture}[scale=0.65]
  \begin{axis}[legend columns=2,
               xmin = 200,
               xmax = 10000,
               grid = major,
               ymode= log,
               log basis y=10,
               ymin = 2e-8,
               ymax = 1,
               ytick = {1,1e-1,1e-3,1e-5,1e-7},
               legend style={at={(0.79,0.63)},anchor=south, font=\footnotesize},
               mark size=2.5pt,
               xlabel={code length $n$},
               ylabel={$\mathtt{DFR}^{(2)}$},
               line width=0.5pt,
               cycle list name=waterfall-floor,
               xticklabel style={
                 /pgf/number format/fixed,
                 /pgf/number format/precision=5
               },
scaled x ticks=false]
   \foreach \vvalue in {7,9,11,15,19} {
     \addplot table [x expr = \thisrow{r}*2,
                     y expr= \thisrow{dfr},
                     restrict y to domain=-18.1:1, 
                     col sep = comma] {data/density_sweep/model/DFR_n0_2_v_\vvalue_t_18.csv};
     \addlegendentryexpanded{};
     \addplot table [x expr = \thisrow{r}*2,
                     y expr = \thisrow{dfr}, 
                     restrict y to domain=-18.3:1, 
                     col sep = comma] {data/density_sweep/sim10e8/DFR_n0_2_v_\vvalue_t_18.csv};
    \addlegendentryexpanded{$v=\vvalue$};
   }

\end{axis}
\end{tikzpicture}

%% file: figures/tikz_pic_density_comparison.tex
\pgfkeys{/pgf/number format/.cd,1000 sep={}}
\begin{tikzpicture}[scale=0.65]
  \begin{axis}[legend columns=2,
               xmin = 200,
               xmax = 12000,
               grid = major,
               ymode= log,
               log basis y=10,
               ymin = 2e-8,
               ymax = 1,
               ytick = {1,1e-1,1e-3,1e-5,1e-7},
               legend style={at={(0.79,0.63)},anchor=south, font=\footnotesize},
               mark size=1.5pt,
               xlabel={code length $n$},
               ylabel={$\mathtt{DFR}^{(2)}$},
               line width=0.5pt,
               cycle list name=waterfall-floor-model,
               xticklabel style={
                 /pgf/number format/fixed,
                 /pgf/number format/precision=5
               },
scaled x ticks=false]
   \foreach \vvalue in {7,9,11,15,19} {
     \addplot table [x expr = \thisrow{r}*2,
                     y expr= \thisrow{dfr},
                     restrict y to domain=-18.1:1, 
                     col sep = comma] {data/density_sweep/model/DFR_n0_2_v_\vvalue_t_18.csv};
     \addlegendentryexpanded{};
     \addplot table [x expr = \thisrow{r}*2,
                     y expr = \thisrow{dfr}, 
                     restrict y to domain=-18.3:1, 
                     col sep = comma] {data/density_sweep_fixed/DFR_n0_2_v_\vvalue_t_18.csv};
    \addlegendentryexpanded{$v=\vvalue$};
   }

\end{axis}
\end{tikzpicture}

%% file: figures/tikz_pic_exhaustive_comparison.tex
\pgfkeys{/pgf/number format/.cd,1000 sep={}}
\begin{tikzpicture}[scale=0.65]
  \begin{axis}[
               legend columns=1,
               xmin = 500,
               xmax = 5000,
               grid = major,
               ymode= log,
               log basis y=10,
               ymin = 2e-6,
               ymax = 1,
               ytick = {1,1e-1,1e-3,1e-5,1e-7},
               legend style={at={(0.80,0.73)},anchor=south, font=\footnotesize},
               mark size=3pt,
               xlabel={code length $n$},
               ylabel={$\mathtt{DFR}^{(2)}$},
               line width=0.5pt,
                mark options=solid,
               xticklabel style={
                 /pgf/number format/fixed,
                 /pgf/number format/precision=5
               },
scaled x ticks=false]

    \addplot[blue, densely dotted, mark=x]
             table [x expr = 2*\thisrow{r},
                    y expr = \thisrow{dfr}, 
                    col sep = comma]{data/exhaustive_th_sweep/DFR_fixed_n0_2_v_19_t_18.csv};
    \addlegendentry{Majority};

    \addplot[black, densely dotted, mark=o]
             table [x expr = 2*\thisrow{r},
                    y expr = \thisrow{dfr}, 
                    col sep = comma]{data/exhaustive_th_sweep/DFR_exhaustive_n0_2_v_19_t_18.csv};
    \addlegendentry{Data driven};

    \addplot[red, densely dotted, mark=x]
             table [x expr = 2*\thisrow{r},
                    y expr = \thisrow{dfr}, 
                    col sep = comma]{data/density_sweep/sim10e8/DFR_n0_2_v_19_t_18.csv};
    \addlegendentry{Our Model};

\end{axis}
\end{tikzpicture}

%% file: figures/tikz_pic_bike_avg_err.tex
\begin{tikzpicture}[scale=1]
  \begin{axis}[
               scale=0.85,
               xmin = 11323,
               xmax = 13323,
               grid = major,
               yminorticks=true,
               ymode= log,
               ymax = 5e4,
               ymin = 2e-5,
               ytick={1e4,1e3,
                      1e2,1e1,
                      1e0,1e-1,
                      1e-2,1e-3,
                      1e-4,1e-5,1e-6},
               minor ytick={9e4 ,8e4 ,7e4 ,6e4 ,5e4 ,4e4 ,3e4 ,2e4 ,1e4 , 
                            9e3 ,8e3 ,7e3 ,6e3 ,5e3 ,4e3 ,3e3 ,2e3 ,1e3 ,
                            9e2 ,8e2 ,7e2 ,6e2 ,5e2 ,4e2 ,3e2 ,2e2 ,1e2 ,
                            9e1 ,8e1 ,7e1 ,6e1 ,5e1 ,4e1 ,3e1 ,2e1 ,1e1 ,
                            9e0 ,8e0 ,7e0 ,6e0 ,5e0 ,4e0 ,3e0 ,2e0 ,1e0 ,
                            9e-1,8e-1,7e-1,6e-1,5e-1,4e-1,3e-1,2e-1,1e-1,
                            9e-2,8e-2,7e-2,6e-2,5e-2,4e-2,3e-2,2e-2,1e-2,
                            9e-3,8e-3,7e-3,6e-3,5e-3,4e-3,3e-3,2e-3,1e-3,
                            9e-4,8e-4,7e-4,6e-4,5e-4,4e-4,3e-4,2e-4,1e-4,
                            9e-5,8e-5,7e-5,6e-5,5e-5,4e-5,3e-5,2e-5,1e-5},
               xtick={11323,11823,12323,12823,13323},
               xticklabel style={
                 /pgf/number format/fixed,
                 /pgf/number format/precision=5},
               scaled x ticks=false,  
               legend style={at={(0.25,0.04)},anchor=south, font=\footnotesize},
               mark size=2pt,
               xlabel={$r$},
               ylabel={Avg. num. of discrepancies},
               line width=0.2pt,
               mark options=solid]

    \addplot[green,mark=x]
             table [x expr = \thisrow{r},
                    y expr = \thisrow{avg_err}, 
                    col sep = comma]{data/bike_sweep_cat1/fixed_th_39_10e6.csv};
    \addlegendentry{Majority + $\delta$};
    \addplot[red,mark=x]
             table [x expr = \thisrow{r},
                    y expr = \thisrow{avg_err}, 
                    col sep = comma]{data/bike_sweep_cat1/bike_th_10e6.csv};
    \addlegendentry{BIKE-flip~\cite{BIKE}};
    \addplot[blue, mark=x]
             table [x expr = \thisrow{r},
                    y expr = \thisrow{avg_err}, 
                    col sep = comma]{data/bike_sweep_cat1/model_th_10e6.csv};
    \addlegendentry{Our approach};
    
    \addplot[ green, mark=*, 
    mark options={scale=1, fill=green}
    ] coordinates {(12323, 3.69503731e+02)};

    \addplot[ red, mark=*, 
    mark options={scale=1, fill=red}
    ] coordinates {(12323, 1.72171369e+01)};

    \addplot[ blue, mark=*, 
    mark options={scale=1, fill=blue}
    ] coordinates {(12323, 6.39280000e-02)};

\end{axis}
\end{tikzpicture}

%% file: appendix0.tex
We begin by defining the transition probabilities for the step
$\left(\mathcal{Z}_0,\mathcal{Z}_1\right)_{(l_{01},l_{11})} \to
\left(\mathcal{Z}_0,\mathcal{Z}_1\right)_{(l_{01}+1,l_{11})}$.
To this end, we denote as $\mathcal{F} \in \{0,1,\ldots,\min(t,w)\}$ the number of
asserted bit in $e$ appearing in a parity-check,
as $\mathcal{F}_{01} \in \{0,1,\ldots,\min(\epsilon_{01},w)\}$ the number of positions in $\jset{0}{1}$ appearing in a parity-check, and as $\mathcal{F}_{11} \in \{0,1,\ldots,\min(\epsilon_{11},w)\}$ the number of positions in $\jset{1}{1}$ appearing in a parity-check. Given a specific parity-check $H_{i,:}$, we now assume to know the values of $\mathcal{F}$, $\mathcal{F}_{01}$ and $\mathcal{F}_{11}$ (say $f$, $f_{01}$ and $f_{11}$), and compute the probability that during the step $\left(\mathcal{Z}_0,\mathcal{Z}_1\right)_{(l_{01},l_{11})} \to \left(\mathcal{Z}_0,\mathcal{Z}_1\right)_{(l_{01}+1,l_{11})}$ the position of the column of $H$ added to the syndrome is included in the $i$-th check, indeed causing a bit flip in the syndrome in position $i$. This event corresponds to selecting a bit in $\jset{0}{1}$ among the $\epsilon_{01}-l_{01}$ not yet selected that is also one of the $f_{01}$ bits in the $i$-th parity-check. The probability that, at the step $(l_{01},l_{11})$, a certain number of bit in $\jset{0}{1}$ already selected $0 \leq b_{01} \leq \min(l_{01},f_{01})$ is included in the $i$-th parity-check is
$$
\Phi_{01}(b_{01}|f_{01},\epsilon_{01}) =
\frac{\binom{f_{01}}{b_{01}} \binom{\epsilon_{01}-f_{01}}{l_{01} - b_{01}}}{\binom{\epsilon_{01}}{l_{01}}}
$$
while the probability of choosing another bit in $\jset{0}{1}$ included in the p.c. during the step $(l_{01}+1,l_{11})$, given $b_{01}$, is
$$
\frac{f_{01} - b_{01}}{\epsilon_{01} - l_{01}}
$$
Additionally, the probability that, at the step $(l_{01},l_{11})$, a certain number of bit in $\jset{1}{1}$ already selected $0 \leq b_{11} \leq \min(l_{11},f_{11})$ is included in the $i$-th parity-check is
$$
\Phi_{11}(b_{11}|f_{11},\epsilon_{11}) =
\frac{\binom{f_{11}}{b_{11}} \binom{\epsilon_{11}-f_{11}}{l_{11} - b_{11}}}{\binom{\epsilon_{11}}{l_{11}}}
$$
We are now interested in computing the probability distribution of $\mathcal{F}_{01}$ and $\mathcal{F}_{11}$, as they are strictly correlated to the value of $\mathcal{F}$, starting from the fact that $f_{01} \leq w-f$ and $f_{11} \leq f$ due to the nature of the bits included in the parity-check. We remind that all the probabilities and p.m.f. are implicitly conditioned to $y$, and that the p.m.f. of $\mathcal{F}$ conditioned to $y$ has already been derived in~\cite{DBLP:conf/isit/AnnechiniBP24}. Since the probability that a parity-check includes $f_{01}$ positions in $\jset{0}{1}$ (or $f_{11}$ positions in $\jset{1}{1}$, respectively) strongly depends on the \textit{parity} of $f$, defining the satisfaction status of the parity-check during the first decoding iteration, we will subdivide the analysis of the p.m.f. of $\mathcal{F}_{01}$ and $\mathcal{F}_{11}$ conditioned to $\mathcal{F}$ in the cases where $f$, the value assumed by $\mathcal{F}$ in a specific parity-check, is either even or odd.

Starting from the case where $f$ is even, we have that the probability of erroneously flipping up a correct bit (i.e. a zero bit in $e$) is $\pflipz{\mathtt{oneEqSat}}$
We define $\eta(f,f_{01})$ as the probability of flipping up $f_{01}$ bits among the $w-f$ correct bits the p.c. during the first iteration:
$
\eta(f,f_{01}) = \bindist{w-f}{\pflipz{\mathtt{oneEqSat}}}{f_{01}}
$
Additionally, we define $\zeta(f,f_{01},\epsilon_{01})$ as the probability of flipping up $\epsilon_{01} - f_{01}$ bits among the $n-t-(w-f)$ correct bits not included in the p.c. during the first iteration:
$
\zeta(f,f_{01},\epsilon_{01}) = \bindist{n-t-(w-f)}{\pflipzstd}{\epsilon_{01} - f_{01}}
$
Since the two event described by $\eta$ and $\zeta$ act on two disjoint and independent sets of positions, we can compute the probability of having $\epsilon_{01}$ incorrect flip ups among the two set of bits as:
$$
\xi(f,\epsilon_{01}) = \sum_{f_{01}=
\max(0,\epsilon_{01}-(n-t-(w-f)))}^{
\min(\epsilon_{01},w-f)
}
\eta(f,f_{01}) \cdot \zeta(f,f_{01},\epsilon_{01})
$$
Finally, we can derive the probability of having $f_{01}$ incorrect flip ups in the p.c., given the value $f$ and the overall number of incorrect flip ups $\epsilon_{01}$ as:
$$
P(\mathcal{F}_{01} = f_{01} | \mathcal{F}=f, |\jset{0}{1}|= \epsilon_{01}) =
$$
$$
 = \psi_{01}(f_{01}|f,\epsilon_{01}) = \frac{\eta(f,f_{01}) \cdot \zeta(f,f_{01},\epsilon_{01})}{\xi(f,\epsilon_{01})}
$$

The probability of maintaining an incorrect bit (i.e. an asserted bit in $e$) included in the parity-check, assuming $f$ to be even, is $\pnoflipo{\mathtt{oneEqSat}}$. We define $\nu(f,f_{11})$ as the probability of flipping up $f_{11}$ bits among the $f$ correct bits the p.c. during the first iteration:
$
\nu(f,f_{11}) = \bindist{f}{\pnoflipo{\mathtt{oneEqSat}}}{f_{11}}
$
Additionally, we define $\lambda(f,f_{11},\epsilon_{11})$ as the probability of maintaining $\epsilon_{11} - f_{11}$ bits among the $t-f$ incorrect bits not included in the p.c. during the first iteration:
$
\lambda(f,f_{11},\epsilon_{11}) = \bindist{t-f}{\pnoflipostd}{\epsilon_{11} - f_{11}}
$
Since the two event described by $\nu$ and $\lambda$ act on two disjoint and independent sets of positions, we can compute the probability of having $\epsilon_{11}$ incorrect bits maintained among the two set of bits as:
$$
\theta(f,\epsilon_{11}) = \sum_{f_{11}=
\max(0,\epsilon_{11}-(t-f))}^{
\min(\epsilon_{11},f)
}
\nu(f,f_{11}) \cdot \lambda(f,f_{11},\epsilon_{11})
$$
Finally, we can derive the probability of having $f_{11}$ incorrect bits maintained in the p.c., given the value $f$ and the overall number of incorrect bits maintained $\epsilon_{11}$ as:
$$
P(\mathcal{F}_{11} = f_{11} | \mathcal{F}=f, |\jset{1}{1}|= \epsilon_{11}) =
$$
$$
 = \psi_{11}(f_{11}|f,\epsilon_{11}) = \frac{\nu(f,f_{11}) \cdot \lambda(f,f_{11},\epsilon_{11})}{\theta(f,\epsilon_{11})}
$$

The derivation of $\psi_{01}(f_{01}|f,\epsilon_{01})$ and $\psi_{11}(f_{11}|f,\epsilon_{11})$ assuming $f$ to be odd instead of even only requires the substitution of $\pflipz{\mathtt{oneEqSat}}$ with $\pflipz{\mathtt{oneEqUnsat}}$, and $\pnoflipo{\mathtt{oneEqSat}}$ with $\pnoflipo{\mathtt{oneEqUnsat}}$.

We now have all the tools required to compute the transition probabilities defining of the Markov process, $\Pr\left(\left(\mathcal{Z}_0 = x_0, \mathcal{Z}_1 = x_1\right)_{(l_{01}+1,l_{11})} | \left(\mathcal{Z}_0 = z_0, \mathcal{Z}_1 = z_1\right)_{(l_{01},l_{11})} \right)$, modeling the state change caused by an addition of a column of $H$ in a position not yet selected in $\jset{0}{1}$. We begin by noting that all the $r-y$ parity-checks that are satisfied before the first iteration (out of which $z_0$ of those are unsatisfied at the current step) have an even value for $\mathcal{F}$, while all the $y$ parity-checks that are satisfied before the first iteration (out of which $z_1$ of those are unsatisfied at the current step) have an odd value for $\mathcal{F}$. Our analysis is therefore structured as follows:
we compute the probability distribution of the number of flips affecting either $\mathcal{Z}_0$ or $\mathcal{Z}_1$ during the current step; we derive the probability of flipping syndrome bits based on their satisfaction both before the first iteration and during the current step (yielding a total of four possibilities); we finally derive the joint p.m.f. of $\left(\left(\mathcal{Z}_0 = x_0, \mathcal{Z}_1 = x_1\right)_{(l_{01}+1,l_{11})} | \left(\mathcal{Z}_0 = z_0, \mathcal{Z}_1 = z_1\right)_{(l_{01},l_{11})} \right)$, completely describing the state change during the step $(l_{01},l_{11}) \to (l_{01}+1,l_{11})$.

The process of adding one column of the parity-check matrix $H$, regardless of the chosen position, will cause a total of $v$ bit flips in the syndrome vector, due to the regularity of $H$. What we want to compute is the probability that, out of $v$ bit flips happening on the syndrome, $a$ of these happen on syndrome bits that were satisfied before the first iteration (thus affecting $\mathcal{Z}_0$), while $v-a$ of these happen on syndrome bits that were unsatisfied before the first iteration (thus affecting $\mathcal{Z}_1$). To this end, we underline that the chosen position is in $\jset{0}{1}$, meaning that its upc is greater than the flipping threshold $\thfirstit$. Since the upc is the number of unsatisfied parity-checks in which the selected bit appears, it is also the number of flips affecting $\mathcal{Z}_1$ during the Markov process. Denoting as $\upcfirstit{i}$ the upc of the selected bit in position $i$ during the first decoding iteration, we can therefore derive the probability of flipping $a$ syndrome bits that were satisfied before the first iteration, and $v-a$ syndrome bits that were unsatisfied before the first iteration, as:
$$
\Pr(\upcfirstit{i} = v-a | i \in \jset{0}{1}) =
$$
$$
=
\begin{cases}
    \frac{\bindist{v}{\punsatz}{v-a}}{\pflipzstd} & \text{if } v-a \geq \thfirstit \\
    0 & \text{otherwise}
\end{cases}
$$
We now proceed deriving the probability of flipping a syndrome bit, given its value (satisfied or unsatisfied) before the first iteration and during the current process. We start from the case where a parity-check is satisfied in both cases.

A parity-check being satisfied before the first iteration implies that its value of $\mathcal{F}$ is even. Additionally, we know that an odd number of bits in either $\jset{0}{1}$ and $\jset{1}{1}$, among the $l_{01}$ and $l_{11}$ already selected, are included in the parity-check. The current satisfaction of a parity-check is thus not defined by the parity of $f_{01} + f_{11}$ (instead defining the satisfaction of the check at the end of the process), but from the fraction of the $f_{01}$ positions in $\jset{0}{1}$ and $f_{11}$ positions in $\jset{1}{1}$ that are both included in the p.c. and selected during the previous steps, denoted as $b_{01}$ and $b_{11}$ respectively. A currently satisfied parity-check is characterized by an even value for $b_{01} + b_{11}$.

The probability $\Pr(\mathcal{F}=f)$ has been derived in~\cite{DBLP:conf/isit/AnnechiniBP24}, while the p.m.f.s of $\mathcal{F}_{01}$ and $\mathcal{F}_{11}$ are fully described by $\psi_{01}(f_{01}|f,\epsilon_{01})$ and $\psi_{11}(f_{11}|f,\epsilon_{11})$, derived before.
Similarly, the probability that the number of positions in $\jset{0}{1}$ and $\jset{1}{1}$ that are both included in the p.c. and selected during the previous steps is equal to $b_{01}$ and $b_{11}$, given the value of $f$, $f_{01}$ and $f_{11}$, have been defined as $\Phi_{01}(b_{01}|f_{01},\epsilon_{01})$ and $\Phi_{11}(b_{11}|f_{11},\epsilon_{11})$. Indicating as $\mathtt{E}_{(s,s)}$ the event of a parity-check being satisfied both before the first iteration and during the current step, we can derive the probability of flipping such parity-check during the step $(l_{01},l_{11}) \to (l_{01}+1,l_{11})$ as:
$$
\pflipsyn{s}{s} = \Pr(\text{flip} | \mathtt{E}_{(s,s)}) =
\frac{\Pr(\text{flip} \cap \mathtt{E}_{(s,s)})}{\Pr(\mathtt{E}_{(s,s)})}
$$
The probability $\Pr(\mathtt{E}_{(s,s)})$ can be derived as the probability that both $f$ and $b_{01}+b_{11}$ are even:
$$
\Pr(\mathtt{E}_{(s,s)}) =
\sum_{f\text{ even}}
\Pr(\mathcal{F}=f) \cdot
$$
$$
\cdot \sum_{f_{01}}
\sum_{f_{11}}
\psi_{01}(f_{01}|f,\epsilon_{01}) \cdot
\psi_{11}(f_{11}|f,\epsilon_{11}) \cdot
$$
$$
\cdot \sum_{b_{01} + b_{11} \text{ even}}
\Phi_{01}(b_{01}|f_{01},\epsilon_{01}) \cdot
\Phi_{11}(b_{11}|f_{11},\epsilon_{11})
$$
The probability $\Pr(\text{flip} \cap \mathtt{E}_{(s,s)})$ can be computed in a similar way, modeling the event that $f$ and $b_{01}+b_{11}$ are even, and that the position in $\jset{0}{1}$ selected during the current step is one of the $f_{01} - b_{01}$ bits in $\jset{0}{1}$ included in the p.c. and not yet selected:
$$
\Pr(\text{flip} \cap \mathtt{E}_{(s,s)}) =
\sum_{f\text{ even}}
\Pr(\mathcal{F}=f) \cdot
$$
$$
\cdot \sum_{f_{01}}
\sum_{f_{11}}
\psi_{01}(f_{01}|f,\epsilon_{01}) \cdot
\psi_{11}(f_{11}|f,\epsilon_{11}) \cdot
$$
$$
\cdot
\sum_{b_{01}}
\sum_{b_{11} |
b_{01} + b_{11} \text{ even}}
\Phi_{01}(b_{01}|f_{01},\epsilon_{01}) \cdot
\Phi_{11}(b_{11}|f_{11},\epsilon_{11}) \cdot \frac{f_{01}-b_{01}}{\epsilon_{01}-l_{01}}
$$

The derivation of $\pflipsyn{s}{u}$, $\pflipsyn{u}{s}$ and $\pflipsyn{u}{u}$, the other probabilities of flipping syndrome bits based on their satisfaction ($s$ or $u$) before the first iteration (first index) and during the current step (second index) is equivalent to the one of $\pflipsyn{s}{s}$, with the only difference being the values of $f$ and $b_{01}+b_{11}$. In particular, the sums regarding $f$ happen over even values if the p.c. is satisfied before the first iteration (in the probabilities having $s$ as first index) and over odd values otherwise (in the probabilities having $u$ as first index); the sums regarding $b_{01} + b_{11}$ happen over even values if the p.c. is satisfied at the current step (in the probabilities having $s$ as second index) and over odd values otherwise (in the probabilities having $u$ as second index).

We now derive the probability that, assuming $a$ flips happening on syndrome bits that were satisfied before the first iteration (thus affecting $\mathcal{Z}_0$) and $v-a$ flips happening on syndrome bits that were unsatisfied before the first iteration (thus affecting $\mathcal{Z}_1$), a certain number of them happen on currently satisfied or unsatisfied checks, thus characterizing the actual change in the values of $\mathcal{Z}_0$ and $\mathcal{Z}_1$. Starting from the $a$ flips affecting $\mathcal{Z}_0$, we know the probability of flipping currently satisfied checks ($\pflipsyn{s}{s}$) and currently unsatisfied checks ($\pflipsyn{s}{u}$). Given the value of $\mathcal{Z}_0$, say $z_0$, we have $r-y-z_0$ currently satisfied checks and $z_0$ currently unsatisfied checks. The probability of $a_{(s,s)}$ flips happening on currently satisfied syndrome bits is
$
\chi_{(s,s)}(a_{(s,s)},z_0) = \bindist{r-y-z_0}{\pflipsyn{s}{s}}{a_{(s,s)}}
$,
while the probability of $a - a_{(s,s)}$ flips happening on currently unsatisfied syndrome bits is
$
\chi_{(s,u)}(a_{(s,s)},a,z_0) = \bindist{z_0}{\pflipsyn{s}{u}}{a - a_{(s,s)}}
$.
Being the two set of bits disjoint and independent, the probability of a total of $a$ flips happening among the two set of bits is
$\chi_{(s,\times)}(a, z_0) =$
$$
\sum_{a_{(s,s)=\max(0,a - z_0)}}^{\max(a, r-y-z_0)}
\chi_{(s,s)}(a_{(s,s)},z_0) \cdot \chi_{(s,u)}(a_{(s,s)},a,z_0)
$$
Finally, the probability that, out of $a$ bit flips, $a_{(s,s)}$ of them happen on currently satisfied bits is
$$
\kappa_{s}(a_{(s,s)},a,z_0) =
\frac{
\chi_{(s,s)}(a_{(s,s)},z_0) \chi_{(s,u)}(a_{(s,s)},a,z_0)
}{
\chi_{(s,\times)}(a, z_0)
}
$$
We now move to the $v-a$ flips affecting $\mathcal{Z}_1$. Given the value of $\mathcal{Z}_1$, say $z_1$, we have $y-z_1$ currently satisfied checks and $z_1$ currently unsatisfied checks. The probability of $a_{u,s}$ flips happening on currently satisfied syndrome bits is
$
\chi_{(u,s)}(a_{(u,s)},z_1) = \bindist{y-z_1}{\pflipsyn{u}{s}}{a_{(u,s)}}
$,
while the probability of $v - a - a_{(u,s)}$ flips happening on currently unsatisfied syndrome bits is
$
\chi_{(u,u)}(a_{(u,s)},a,z_1) = \bindist{z_1}{\pflipsyn{u}{u}}{v - a - a_{(u,s)}}
$.
Being the two set of bits disjoint and independent, the probability of a total of $v-a$ flips happening among the two set of bits is
$
\chi_{(u,\times)}(a, z_1) =
$
$$
\sum_{a_{(u,s)=\max(0,v - a - z_1)}}^{\max(v-a, y-z_1)}
\chi_{(u,s)}(a_{(u,s)},z_1) \cdot \chi_{(u,u)}(a_{(u,s)},a,z_1)
$$
Finally, the probability that, out of $v-a$ bit flips, $a_{(u,s)}$ of them happen on currently satisfied bits is
$$
\kappa_{u}(a_{(u,s)},a,z_1) =
\frac{
\chi_{(u,s)}(a_{(u,s)},z_1) \chi_{(u,u)}(a_{(u,s)},a,z_1)
}{
\chi_{(u,\times)}(a, z_1)
}
$$
The derivation of $\kappa_s$ and $\kappa_u$ allows us to explicitly compute
$\Pr\left(\left(\mathcal{Z}_0 = x_0, \mathcal{Z}_1 = x_1\right)_{(l_{01}+1,l_{11})} | \left(\mathcal{Z}_0 = z_0, \mathcal{Z}_1 = z_1\right)_{(l_{01},l_{11})} \right)$, since we now have both the probability distribution of the number of flips affecting $\mathcal{Z}_0$ and $\mathcal{Z}_1$, and the fraction of each flip happening on currently satisfied or unsatisfied checks.
We begin by noting that, in the case where $a$ flips affect syndrome bits that were satisfied before the first iteration, $a_{(s,s)}$ flips affect bits that were satisfied before the first iteration and are currently satisfied, $a_{(u,s)}$ flips affect bits that were unsatisfied before the first iteration and are currently satisfied, then the resulting state after the bit flips is $\left(\mathcal{Z}_0 = x_0, \mathcal{Z}_1 = x_1\right)_{(l_{01}+1,l_{11})}$ with $x_0 = z_0 - a + 2a_{(s,s)}$, $x_1 = z_1 - (v-a) + 2a_{(u,s)}$: from this relation, we get $a_{(s,s)} = \frac{x_0 - z_0 + a}{2}$, $a_{(u,s)} = \frac{x_1 - z_1 + (v-a)}{2}$. To obtain the state change probability, we need to average the previously found terms over all possible values of $a$:
$$
\Pr\left(\left(\mathcal{Z}_0 = x_0, \mathcal{Z}_1 = x_1\right)_{(l_{01}+1,l_{11})} | \left(\mathcal{Z}_0 = z_0, \mathcal{Z}_1 = z_1\right)_{(l_{01},l_{11})} \right)
$$
$$
=
\sum_{a=0}^{v}
\Pr(\upcfirstit{i} = v-a | i \in \jset{0}{1}) \cdot
$$
$$
\cdot \kappa_{s}\left(\frac{x_0 - z_0 + a}{2},a,z_0\right)
\cdot \kappa_{u}\left(\frac{x_1 - z_1 + (v-a)}{2},a,z_1\right)
$$

The explicit derivation of the joint p.m.f. of $\left(\left(\mathcal{Z}_0 = x_0, \mathcal{Z}_1 = x_1\right)_{(l_{01},l_{11}+1)} | \left(\mathcal{Z}_0 = z_0, \mathcal{Z}_1 = z_1\right)_{(l_{01},l_{11})} \right)$, associated to the step $(l_{01},l_{11}) \to (l_{01},l_{11}+1)$ where a bit in $\jset{1}{1}$ is selected instead of a bit in $\jset{0}{1}$, is almost equivalent to the one reported previously. The first difference is in the formula of $\Pr(\text{flip} \cap \mathtt{E}_{(s,s)})$, used in the derivation of $\pflipsyn{s}{s}$ (the same modification applies for $\pflipsyn{s}{u}$, $\pflipsyn{u}{s}$ and $\pflipsyn{u}{u}$), where the probability of flipping the satisfaction state of a parity-check given $f$, $f_{11}$ and $b_{11}$ is
$$
\frac{f_{11} - b_{11}}{\epsilon_{11} - l_{11}}
$$
due to the fact that the selected bit is in $\jset{1}{1}$. Moreover, the term $\Pr(\upcfirstit{i} = v-a | i \in \jset{0}{1})$ in the formula for the state change probability is replaced by $\Pr(\upcfirstit{i} = v-a | i \in \jset{1}{1})$, due to the different nature of the selected bit:
$$
\Pr(\upcfirstit{i} = v-a | i \in \jset{1}{1}) =
$$
$$
=
\begin{cases}
    \frac{\bindist{v}{\punsato}{v-a}}{\pnoflipostd} & \text{if } v-a \leq \thfirstit-1 \\
    0 & \text{otherwise}
\end{cases}
$$

%% file: appendix2.tex
In this section, we derive the probability of flipping a specific bit during the second iteration of the parallel decoder, depending on its correctness before and after the first iteration. We denote such probabilities as $\pflipzz$ for bits in $\jset{0}{0}$, $\pflipzo$ for bits in $\jset{0}{1}$, $\pflipoz$ for bits in $\jset{1}{0}$ and $\pflipoo$ for bits in $\jset{1}{1}$.

We begin by computing the probability $\pflipzz$ of incorrectly flipping a bit in $\jset{0}{0}$, containing all the correct bits ($e_i = 0$) that have been correctly maintained during the first iteration ($\discvectoridx{i}=0$). To this end, we are interested in deriving the probability $\pcheck{00}{s}{u}$ that a satisfied parity check, where a bit in $\jset{0}{0}$ appears, becomes unsatisfied after the first iteration. We define $\sevent{s}$ as the event that, choosing an asserted bit in $H$ in position $(i,j)$ (so that $h_{i,j}=1$), the $i$-th parity check is satisfied before the first iteration, and as $\sevent{s \to u}$ the event of said parity check being satisfied before the first iteration and unsatisfied after the first iteration. Likewise, we denote as $\sevent{\jset{0}{0}}$ the event of choosing a position $j \in \jset{0}{0}$ among the $n$ positions of the error vector. 
Finally, we define $\mathcal{F}$ as the random variable corresponding to the number of
asserted bit in $e$ appearing in a parity-check,
$\mathcal{F}_{01}$ as the random variable corresponding to the number of positions in $\jset{0}{1}$ appearing in a parity-check, $\mathcal{F}_{11}$ as the random variable corresponding to the number of positions in $\jset{1}{1}$ appearing in a parity-check. The probability $\pcheck{00}{s}{u}$ can be derived as
$$
\pcheck{00}{s}{u} = \condprob{\sevent{s \to u}}{\sevent{s} \cap \sevent{\jset{0}{0}}} =
$$
$$
=
\frac{\Pr\left(\sevent{s \to u} \cap \sevent{s} \cap \sevent{\jset{0}{0}}\right)}{\Pr\left( \sevent{s} \cap \sevent{\jset{0}{0}}\right)} =
$$
$$
=
\frac{\Pr\left(\sevent{s \to u} \cap \sevent{\jset{0}{0}} \right)}{\Pr\left(  \sevent{s} \cap \sevent{\jset{0}{0}}\right)}
$$
where the last step is justified by the fact that $\sevent{s} \cap \sevent{s \to u} = \sevent{s \to u}$. Given that the satisfaction of a parity check is determined by the parity of $\mathcal{F}$ during the first iteration and by the parity of $\mathcal{F}_{01}+ \mathcal{F}_{11}$ during the second iteration, the term $\Pr\left(\sevent{s \to u} \cap \sevent{\jset{0}{0}} \right)$ corresponds to the probability that the value of $\mathcal{F}$ in the parity check (say, $f$) is even, that the sum of the values of $\mathcal{F}_{01}$ and $\mathcal{F}_{11}$ in the parity check (say, $f_{01}$ and $f_{11}$) is odd, and that a bit in $\jset{0}{0}$ is selected among the $w-f-f_{01}$ positions in $\jset{0}{0}$ included in the parity check. The probability $\Pr\left(\sevent{s \to u} \cap \sevent{\jset{0}{0}} \right)$ can thus be derived as
$$
\sum_{
f \text{ even}} 
\sum_{
f_{01}}
\sum_{
\substack{f_{11} \\ f_{01}+f_{11} \text{ odd}}
}
\Pr(f,f_{01},f_{11}|\epsilon_{01},\epsilon_{11},z_0,z_1)
\cdot \frac{w-f-f_{01}}{w}
$$
where $\Pr(f,f_{01},f_{11}|\epsilon_{01},\epsilon_{11},z_0,z_1)$ denotes the joint p.m.f. of $(\mathcal{F},\mathcal{F}_{01},\mathcal{F}_{11})$ conditioned to the syndrome status after the first iteration, identified by $z_0$ and $z_1$. The derivation of $\Pr(f,f_{01},f_{11}|\epsilon_{01},\epsilon_{11},z_0,z_1)$ can be found in Appendix \ref{app:distrf}.
The probability $\Pr\left(  \sevent{s} \cap \sevent{\jset{0}{0}}\right)$ can be derived in a similar fashion by noting that the parity check is only assumed to be satisfied before the first iteration, thus removing the constraint on the parity of $f_{01}+f_{11}$:
$$
\sum_{f \text{ even}} 
\sum_{f_{01}}
\sum_{f_{11}}
\Pr(f,f_{01},f_{11}|\epsilon_{01},\epsilon_{11},z_0,z_1)
\cdot \frac{w-f-f_{01}}{w}
$$

We now have all the terms required to compute $\pcheck{00}{s}{u}$. The probability $\pcheck{00}{u}{u}$ that an \textit{unsatisfied} parity check, where a bit in $\jset{0}{0}$ appears,
remains unsatisfied after the first iteration can be computed as
$$
\pcheck{00}{u}{u}
= \frac{\Pr\left(\sevent{u \to u} \cap \sevent{\jset{0}{0}} \right)}{\Pr\left(  \sevent{u} \cap \sevent{\jset{0}{0}}\right)}
$$
The full derivation of $\Pr\left(\sevent{u \to u} \cap \sevent{\jset{0}{0}} \right)$ and $\Pr\left(  \sevent{u} \cap \sevent{\jset{0}{0}}\right)$ is equivalent to the one of $\Pr\left(\sevent{s \to u} \cap \sevent{\jset{0}{0}} \right)$ and $\Pr\left(  \sevent{s} \cap \sevent{\jset{0}{0}}\right)$, with the only difference being that the sum over $f$ ranges over odd values instead of even values.

The probabilities $\pcheck{00}{s}{u}$ and $\pcheck{00}{u}{u}$ allow us to model the behavior of the parity equations where a bit in $\jset{0}{0}$ is included, given their satisfaction before the first iteration. Assuming the initial upc $\upcfirstit{i}$ of a bit $i \in \jset{0}{0}$ to be equal to $\mathtt{u}_1$, in order for its upc after the second iteration $\upcsecondit{i}$ to be equal to $\mathtt{u}_2$, a total of $\mathtt{u}_2$ parity checks among the $v-\mathtt{u}_1$ satisfied ones and $\mathtt{u}_1$ unsatisfied ones must become unsatisfied after the first iteration. The probability of such an event happening, denoted as $\condprob{\upcsecondit{i} = \mathtt{u}_2}{\upcfirstit{i} = \mathtt{u}_1, i \in \jset{0}{0}}$, can be derived as the probability that $a$ satisfied parity checks become unsatisfied and $\mathtt{u}_2 - a$ unsatisfied parity checks remain unsatisfied, for all possible values of $a$:
$$
\sum_{a = \max(0, \mathtt{u}_2 - \mathtt{u}_1)}^{\min(\mathtt{u}_2, v-\mathtt{u}_1)}
\bindist{v - \mathtt{u}_1}{\pcheck{00}{s}{u}}{a} \cdot
\bindist{\mathtt{u}_1}{\pcheck{00}{u}{u}}{\mathtt{u}_2 - a}
$$
Moreover, the p.m.f. of $\upcfirstit{i}$ assuming $i \in \jset{0}{0}$ is
$$
\Pr(\upcfirstit{i} = \mathtt{u}_1 | i \in \jset{0}{0}) =
$$
$$
=
\begin{cases}
    \frac{\bindist{v}{\punsatz}{\mathtt{u}_1}}{1-\pflipzstd} & \text{if } \mathtt{u}_1 < \thfirstit \\
    0 & \text{otherwise}
\end{cases}
$$
due to the fact that the bit in $\jset{0}{0}$ is correct ($e_i = 0$) and has not been flipped during the first iteration, meaning that its upc is lower than the flipping threshold $\thfirstit$.

We can now compute the probability $\pflipzz$ of incorrectly flipping a bit in $\jset{0}{0}$ as the probability that its upc after the second iteration $\upcsecondit{i}$ is greater than the flipping threshold $\thsecondit$:
$$
\pflipzz =
\sum_{\mathtt{u}_2 = \thsecondit}^v
\sum_{\mathtt{u}_1 = 0}^{\thfirstit-1}
\Pr(\upcfirstit{i} = \mathtt{u}_1 | i \in \jset{0}{0}) \cdot
$$
$$
\cdot
\condprob{\upcsecondit{i} = \mathtt{u}_2}{\upcfirstit{i} = \mathtt{u}_1, i \in \jset{0}{0}}
$$

We now move to the derivation of $\pflipzo$, $\pflipoz$ and $\pflipoo$, the probability of flipping bits in $\jset{0}{1}$,  $\jset{1}{0}$ and $\jset{1}{1}$ respectively, highlighting the differences with the formulas for the computation of $\pflipzz$.

Starting from $\pflipzo$, we derive the probability $\pcheck{01}{s}{u}$ that a satisfied parity check, where a bit in $\jset{0}{1}$ appears, becomes unsatisfied after the first iteration with the same line of reasoning as for $\pflipzz$ (changes highlighted in blue):
$$
\pcheck{01}{s}{u} = \condprob{\sevent{s \to u}}{\sevent{s} \cap \sevent{\mathalert{\jset{0}{1}}}} =
$$
$$
=
\frac{\Pr\left(\sevent{s \to u} \cap \sevent{s} \cap \sevent{\mathalert{\jset{0}{1}}}\right)}{\Pr\left( \sevent{s} \cap \sevent{\mathalert{\jset{0}{1}}}\right)} = 
$$
$$
=
\frac{\Pr\left(\sevent{s \to u} \cap \sevent{\mathalert{\jset{0}{1}}} \right)}{\Pr\left(  \sevent{s} \cap \sevent{\mathalert{\jset{0}{1}}}\right)}
$$
The term $\Pr\left(\sevent{s \to u} \cap \sevent{\jset{0}{1}} \right)$ corresponds to the
probability that in the selected equation $f$ is even (being the
parity equation satisfied before the first iteration), $f_{01} + f_{11}$ is
odd (being the parity check unsatisfied after the first iteration),
and that a bit in $\jset{0}{1}$ is selected among the $f_{01}$ positions
in $\jset{0}{1}$ included in the parity check. The probability $\Pr\left(\sevent{s \to u} \cap \sevent{\jset{0}{1}} \right)$ can thus be derived as
$$
\sum_{f \text{ even}}
\sum_{f_{01}}
\sum_{\substack{f_{11} \\ f_{01}+f_{11} \text{ odd}}}
\Pr(f,f_{01},f_{11}|\epsilon_{01},\epsilon_{11},z_0,z_1)
\cdot \mathalert{\frac{f_{01}}{w}}
$$
The probability $\Pr\left( \sevent{s} \cap \sevent{\jset{0}{1}}\right)$ can be derived  removing the constraint
on the parity of $f_{01} + f_{11}$:
$$
\sum_{f \text{ even}}
\sum_{f_{01}}
\sum_{f_{11}}
\Pr(f,f_{01},f_{11}|\epsilon_{01},\epsilon_{11},z_0,z_1)
\cdot \mathalert{\frac{f_{01}}{w}}
$$

The probability $\pcheck{01}{u}{u}$ that an unsatisfied parity check, where a
bit in $\jset{0}{1}$ appears, remains unsatisfied after the first iteration
can be computed as
$$
\pcheck{01}{u}{u} =
\frac{\Pr\left(\sevent{u \to u} \cap \sevent{\mathalert{\jset{0}{1}}} \right)}{\Pr\left(  \sevent{u} \cap \sevent{\mathalert{\jset{0}{1}}}\right)}
$$
The full derivation of $\Pr\left(\sevent{u \to u} \cap \sevent{\jset{0}{1}} \right)$ and $\Pr\left(  \sevent{u} \cap \sevent{\jset{0}{1}}\right)$ is equivalent to the one of
$\Pr\left(\sevent{s \to u} \cap \sevent{\jset{0}{1}} \right)$ and $\Pr\left( \sevent{s} \cap \sevent{\jset{0}{1}}\right)$, with the
only difference being that the sum over $f$ ranges over odd
values instead of even values.

The probability $\condprob{\upcsecondit{i} = \mathtt{u}_2}{\upcfirstit{i} = \mathtt{u}_1, i \in \jset{0}{1}}$ can be derived as the probability that $a$ satisfied parity checks become unsatisfied and $\mathtt{u}_2 - a$ unsatisfied parity checks remain unsatisfied, for all possible values of $a$:
$$ 
\sum_{a = \max(0, \mathtt{u}_2 - \mathtt{u}_1)}^{\min(\mathtt{u}_2, v-\mathtt{u}_1)} 
\bindist{v - \mathtt{u}_1}{\mathalert{\pcheck{01}{s}{u}}}{a} \cdot 
\bindist{\mathtt{u}_1}{\mathalert{\pcheck{01}{u}{u}}}{\mathtt{u}_2 - a}
$$
Moreover, the p.m.f. of $\upcfirstit{i}$ assuming $i \in \jset{0}{1}$ is
$$
\Pr(\upcfirstit{i} = \mathtt{u}_1 | i \in\mathalert{\jset{0}{1}}) =
$$
$$
=
\begin{cases}
    \frac{\bindist{v}{\punsatz}{\mathtt{u}_1}}{\mathalert{\pflipzstd}} & \text{if } \mathalert{\mathtt{u}_1 \geq \thfirstit} \\
    0 & \text{otherwise}
\end{cases}
$$
due to the fact that the bit in $\jset{0}{1}$ is correct ($e_i = 0$) and has been flipped during the first iteration, meaning that its upc is at least equal to the flipping threshold $\thfirstit$.

We can now compute the probability $\pflipzo$ of correctly flipping a bit in $\jset{0}{1}$ as the probability that its upc after the second iteration $\upcsecondit{i}$ is greater than the flipping threshold $\thsecondit$:
$$
\pflipzo = 
\sum_{\mathtt{u}_2 = \thsecondit}^v
\sum_{\mathtt{u}_1 = \mathalert{\thfirstit}}^{\mathalert{v}} 
\Pr(\upcfirstit{i} = \mathtt{u}_1 | i \in \mathalert{\jset{0}{1}}) \cdot
$$
$$
\cdot
\condprob{\upcsecondit{i} = \mathtt{u}_2}{\upcfirstit{i} = \mathtt{u}_1, i \in \mathalert{\jset{0}{1}}}
$$

Moving to $\pflipoz$, we derive the probability $\pcheck{10}{s}{u}$ that a satisfied parity check, where a bit in $\jset{1}{0}$ appears, becomes unsatisfied after the first iteration:
$$
\pcheck{10}{s}{u} = 
\frac{\Pr\left(\sevent{s \to u} \cap \sevent{\mathalert{\jset{1}{0}}} \right)}{\Pr\left(  \sevent{s} \cap \sevent{\mathalert{\jset{1}{0}}}\right)}
$$
The probability $\Pr\left(\sevent{s \to u} \cap \sevent{\jset{1}{0}} \right)$ can be computed as
$$
\sum_{f \text{ even}}
\sum_{f_{01}}
\sum_{\substack{f_{11} \\ f_{01}+f_{11} \text{ odd}}}
\Pr(f,f_{01},f_{11}|\epsilon_{01},\epsilon_{11},z_0,z_1)
\cdot \mathalert{\frac{f-f_{11}}{w}}
$$
The term $\Pr\left( \sevent{s} \cap \sevent{\jset{1}{0}}\right)$ can be derived  removing the constraint
on the parity of $f_{01} + f_{11}$:
$$
\sum_{f \text{ even}}
\sum_{f_{01}}
\sum_{f_{11}}
\Pr(f,f_{01},f_{11}|\epsilon_{01},\epsilon_{11},z_0,z_1)
\cdot \mathalert{\frac{f-f_{11}}{w}}
$$

The probability $\pcheck{10}{u}{u}$ that an unsatisfied parity check, where a
bit in $\jset{1}{0}$ appears, remains unsatisfied after the first iteration
can be computed as
$$
\pcheck{10}{u}{u} =
\frac{\Pr\left(\sevent{u \to u} \cap \sevent{\mathalert{\jset{1}{0}}} \right)}{\Pr\left(  \sevent{u} \cap \sevent{\mathalert{\jset{1}{0}}}\right)}
$$
The full derivation of $\Pr\left(\sevent{u \to u} \cap \sevent{\jset{1}{0}} \right)$ and $\Pr\left(  \sevent{u} \cap \sevent{\jset{1}{0}}\right)$ is equivalent to the one of
$\Pr\left(\sevent{s \to u} \cap \sevent{\jset{1}{0}} \right)$ and $\Pr\left( \sevent{s} \cap \sevent{\jset{1}{0}}\right)$, with the
only difference being that the sum over $f$ ranges over odd
values instead of even values.

The probability $\condprob{\upcsecondit{i} = \mathtt{u}_2}{\upcfirstit{i} = \mathtt{u}_1, i \in \jset{1}{0}}$ can be derived as the probability that $a$ satisfied parity checks become unsatisfied and $\mathtt{u}_2 - a$ unsatisfied parity checks remain unsatisfied, for all possible values of $a$:
$$ 
\sum_{a = \max(0, \mathtt{u}_2 - \mathtt{u}_1)}^{\min(\mathtt{u}_2, v-\mathtt{u}_1)} 
\bindist{v - \mathtt{u}_1}{\mathalert{\pcheck{10}{s}{u}}}{a} \cdot 
\bindist{\mathtt{u}_1}{\mathalert{\pcheck{10}{u}{u}}}{\mathtt{u}_2 - a}
$$
Moreover, the p.m.f. of $\upcfirstit{i}$ assuming $i \in \jset{1}{0}$ is
$$
\Pr(\upcfirstit{i} = \mathtt{u}_1 | i \in \mathalert{\jset{1}{0}}) =
$$
$$
=
\begin{cases}
    \frac{\bindist{v}{\punsato}{\mathtt{u}_1}}{\mathalert{1-\pnoflipostd}} & \text{if } \mathalert{\mathtt{u}_1 \geq \thfirstit} \\
    0 & \text{otherwise}
\end{cases}
$$
due to the fact that the bit in $\jset{1}{0}$ is incorrect ($e_i = 1$) and has been correctly flipped during the first iteration, meaning that its upc is at least equal to the flipping threshold $\thfirstit$.

We can now compute the probability $\pflipoz$ of incorrectly flipping a bit in $\jset{1}{0}$ as the probability that its upc after the second iteration $\upcsecondit{i}$ is greater than the flipping threshold $\thsecondit$:
$$
\pflipoz = 
\sum_{\mathtt{u}_2 = \thsecondit}^v
\sum_{\mathtt{u}_1 = \mathalert{\thfirstit}}^{\mathalert{v}} 
\Pr(\upcfirstit{i} = \mathtt{u}_1 | i \in \mathalert{\jset{1}{0}}) \cdot
$$
$$
\cdot
\condprob{\upcsecondit{i} = \mathtt{u}_2}{\upcfirstit{i} = \mathtt{u}_1, i \in \mathalert{\jset{1}{0}}}
$$

Moving to $\pflipoo$, we derive the probability $\pcheck{11}{s}{u}$ that a satisfied parity check, where a bit in $\jset{1}{1}$ appears, becomes unsatisfied after the first iteration:
$$
\pcheck{11}{s}{u} = 
\frac{\Pr\left(\sevent{s \to u} \cap \sevent{\mathalert{\jset{1}{1}}} \right)}{\Pr\left(  \sevent{s} \cap \sevent{\mathalert{\jset{1}{1}}}\right)}
$$
The probability $\Pr\left(\sevent{s \to u} \cap \sevent{\jset{1}{1}} \right)$ can be computed as
$$
\sum_{f \text{ even}}
\sum_{f_{01}}
\sum_{\substack{f_{11} \\ f_{01}+f_{11} \text{ odd}}}
\Pr(f,f_{01},f_{11}|\epsilon_{01},\epsilon_{11},z_0,z_1)
\cdot \mathalert{\frac{f_{11}}{w}}
$$
The term $\Pr\left( \sevent{s} \cap \sevent{\jset{1}{1}}\right)$ can be derived  removing the constraint
on the parity of $f_{01} + f_{11}$:
$$
\sum_{f \text{ even}}
\sum_{f_{01}}
\sum_{f_{11}}
\Pr(f,f_{01},f_{11}|\epsilon_{01},\epsilon_{11},z_0,z_1)
\cdot \mathalert{\frac{f_{11}}{w}}
$$

The probability $\pcheck{11}{u}{u}$ that an unsatisfied parity check, where a
bit in $\jset{1}{1}$ appears, remains unsatisfied after the first iteration
can be computed as
$$
\pcheck{11}{u}{u} =
\frac{\Pr\left(\sevent{u \to u} \cap \sevent{\mathalert{\jset{1}{1}}} \right)}{\Pr\left(  \sevent{u} \cap \sevent{\mathalert{\jset{1}{1}}}\right)}
$$
The full derivation of $\Pr\left(\sevent{u \to u} \cap \sevent{\jset{1}{1}} \right)$ and $\Pr\left(  \sevent{u} \cap \sevent{\jset{1}{1}}\right)$ is equivalent to the one of
$\Pr\left(\sevent{s \to u} \cap \sevent{\jset{1}{1}} \right)$ and $\Pr\left( \sevent{s} \cap \sevent{\jset{1}{1}}\right)$, with the
only difference being that the sum over $f$ ranges over odd
values instead of even values.

The probability $\condprob{\upcsecondit{i} = \mathtt{u}_2}{\upcfirstit{i} = \mathtt{u}_1, i \in \jset{1}{1}}$ can be derived as the probability that $a$ satisfied parity checks become unsatisfied and $\mathtt{u}_2 - a$ unsatisfied parity checks remain unsatisfied, for all possible values of $a$:
$$ 
\sum_{a = \max(0, \mathtt{u}_2 - \mathtt{u}_1)}^{\min(\mathtt{u}_2, v-\mathtt{u}_1)} 
\bindist{v - \mathtt{u}_1}{\mathalert{\pcheck{11}{s}{u}}}{a} \cdot 
\bindist{\mathtt{u}_1}{\mathalert{\pcheck{11}{u}{u}}}{\mathtt{u}_2 - a}
$$
Moreover, the p.m.f. of $\upcfirstit{i}$ assuming $i \in \jset{1}{1}$ is
$$
\Pr(\upcfirstit{i} = \mathtt{u}_1 | i \in \mathalert{\jset{1}{1}}) =
$$
$$
=
\begin{cases}
    \frac{\bindist{v}{\punsato}{\mathtt{u}_1}}{\mathalert{\pnoflipostd}} & \text{if } \mathtt{u}_1 < \thfirstit \\
    0 & \text{otherwise}
\end{cases}
$$
due to the fact that the bit in $\jset{1}{1}$ is incorrect ($e_i = 1$) and has been incorrectly maintained during the first iteration, meaning that its upc is lower the flipping threshold $\thfirstit$.

We can now compute the probability $\pflipoo$ of correctly flipping a bit in $\jset{1}{1}$ as the probability that its upc after the second iteration $\upcsecondit{i}$ is greater than the flipping threshold $\thsecondit$:
$$
\pflipoo = 
\sum_{\mathtt{u}_2 = \thsecondit}^v
\sum_{\mathtt{u}_1 = 0}^{\mathalert{\thfirstit-1}} 
\Pr(\upcfirstit{i} = \mathtt{u}_1 | i \in \mathalert{\jset{1}{1}}) \cdot
$$
$$
\cdot
\condprob{\upcsecondit{i} = \mathtt{u}_2}{\upcfirstit{i} = \mathtt{u}_1, i \in \mathalert{\jset{1}{1}}}
$$

%% file: appendix1.tex
The aim of this section is to derive the joint p.m.f. of $(\mathcal{F},\mathcal{F}_{01},\mathcal{F}_{11})$, corresponding to the number of incorrect bits in a parity check and bits in $\jset{0}{1}$ and $\jset{1}{1}$ appearing in the p.c., conditioned to the number of positions in $\jset{0}{1}$ and $\jset{1}{1}$, $\mathcal{Z}_0$ (the number of satisfied parity checks becoming unsatisfied) and $\mathcal{Z}_1$ (the number of unsatisfied parity checks remaining unsatisfied), assuming their values to be $\epsilon_{01}$, $\epsilon_{11}$, $z_0$ and $z_1$, respectively. We remind that all the probabilities and p.m.f.s employed in the following are implicitly conditioned to the syndrome weight $y$ before the first iteration. 

Our goal is to compute the following: 
$$
\scriptstyle
\Pr\left((\mathcal{F},\mathcal{F}_{01}, \mathcal{F}_{11}) = (f,f_{01},f_{11}) \mid |\jset{0}{1}|=\epsilon_{01}, |\jset{1}{1}|=\epsilon_{11}, (\mathcal{Z}_0,\mathcal{Z}_1)=(z_0,z_1)\right)
$$
denoted from now on as $\Pr(f,f_{01},f_{11}|\epsilon_{01},\epsilon_{11},z_0,z_1)$ for brevity. We have that:
$$
\Pr(f,f_{01},f_{11}|\epsilon_{01},\epsilon_{11},z_0,z_1) =
$$
$$
= \frac{\Pr(f,f_{01},f_{11},z_0,z_1|\epsilon_{01},\epsilon_{11})}
{\Pr(z_0,z_1|\epsilon_{01},\epsilon_{11})} =
$$
$$
= \frac{\Pr(z_0,z_1|f,\epsilon_{01},\epsilon_{11},f_{01},f_{11}) 
\cdot \Pr(f,f_{01},f_{11}|\epsilon_{01},\epsilon_{11})
}
{\Pr(z_0,z_1|\epsilon_{01},\epsilon_{11})}
$$
The term $\Pr(f,f_{01},f_{11}|\epsilon_{01},\epsilon_{11})$, not depending on the value of $z_0$ and $z_1$, can be expressed as
$$
\Pr(\mathcal{F}=f) 
\cdot \psi_{01}(f_{01}|f,\epsilon_{01})
\cdot \psi_{11}(f_{11}|f,\epsilon_{11})
$$
where $\psi_{01}(f_{01}|f,\epsilon_{01})$ and $\psi_{11}(f_{11}|f,\epsilon_{11})$, the unconstrained p.m.f.s of $\mathcal{F}_{01}$ and $\mathcal{F}_{11}$, have been derived in Appendix \ref{app:trmatrix}. Likewise, $\Pr(z_0,z_1|\epsilon_{01},\epsilon_{11})$ is the unconditional p.m.f. of $(\mathcal{Z}_0,\mathcal{Z}_1)$ already derived in Appendix \ref{app:trmatrix} as
$$
\Pr\left((\mathcal{Z}_0=z_0,\mathcal{Z}_1=z_1)_{(\epsilon_{01},\epsilon_{11})}\right)
$$
We now need to compute $\Pr(z_0,z_1|f,\epsilon_{01},\epsilon_{11},f_{01},f_{11})$ corresponding to the probability of having $z_0$ satisfied parity checks turning unsatisfied and $z_1$ parity checks remaining unsatisfied after the first iteration, given the values $f$, $f_{01}$ and $f_{11}$ of one of the parity checks. To this end, we modify the time-dependent non-homogeneous Markov chain employed in Appendix \ref{app:trmatrix} to derive $\Pr\left((\mathcal{Z}_0=z_0,\mathcal{Z}_1=z_1)_{(\epsilon_{01},\epsilon_{11})}\right)$, in order to include the additional information given by $f$, $f_{01}$ and $f_{11}$. This analysis strongly depends on the parity of $f$ and $f_{01}+f_{11}$, denoting the satisfaction of the p.c. before and after the first iteration. 

Starting from the case where $f$ is even, the hypothesis is that, among the $r-y$ parity equations that are satisfied before the first iteration, one of them undergoes exactly $f_{01}$ flips when performing steps $(l_{01},l_{11}) \to (l_{01}+1,l_{11})$, and exactly $f_{11}$ when performing steps $(l_{01},l_{11}) \to (l_{01},l_{11}+1)$. Indeed, in all the steps where the specific p.c. is flipped, only $v-1$ bit flips happen among the other $r-1$ syndrome bits, while in all the steps where the specific p.c. is \textit{not} flipped, all the $v$ bit flips happen on the other $r-1$ syndrome bits. Additionally, given that the selected parity check is initially satisfied, we know that all the bits included in the p.c. have an upc not greater than $v-1$.
We now proceed, without loss of generality, assuming that all the first $f_{01}$ steps of the type $(l_{01},l_{11}) \to (l_{01}+1,l_{11})$ and all the first $f_{11}$ of the type $(l_{01},l_{11}) \to (l_{01},l_{11}+1)$ flip the selected p.c., while the successive ones do not impact the parity of the selected equation. 

We begin by analyzing the steps $(l_{01},l_{11}) \to (l_{01}+1,l_{11})$ where $l_{01} < f_{01}$. The first difference with respect to the calculations in Appendix \ref{app:trmatrix} is in the derivation of $\chi_{(s,s)}(a_{(s,s)},z_0)$ and $\chi_{(s,u)}(a_{(s,s)},a,z_0)$, denoting the probability that $a_{(s,s)}$ flips happen on parity checks that were satisfied before the first iteration and are currently satisfied and that $a - a_{(s,s)}$ flips happen on parity checks that were satisfied before the first iteration and are currently unsatisfied (for a fixed value $a$). By hypothesis, we assumed that one of the p.c.s satisfied before the first iteration is flipped during the current step (since $l_{01} < f_{01}$). Moreover, said parity check is currently satisfied if $l_{01}+\min(l_{11},f_{11})$ is even, and unsatisfied otherwise. The definition of $\chi_{(s,s)}(a_{(s,s)},z_0)$ and $\chi_{(s,u)}(a_{(s,s)},a,z_0)$, with the inclusion of these information, is $\chi_{(s,s)}(a_{(s,s)},z_0) = $
$$
\scriptscriptstyle
\begin{cases}
    \bindist{r-y-z_0-1}{\pflipsyn{s}{s}}{a_{(s,s)}-1}\\ 
    \quad \quad \quad \text{if $l_{01}+\min(l_{11},f_{11})$ even, $a_{(s,s)}\geq 1$}\\
    0\\
    \quad \quad \quad \text{if $l_{01}+\min(l_{11},f_{11})$ even, $a_{(s,s)}=0$}\\
    \bindist{r-y-z_0}{\pflipsyn{s}{s}}{a_{(s,s)}}\\
    \quad \quad \quad \text{if $l_{01}+\min(l_{11},f_{11})$ odd}\\
\end{cases}
$$
and $\chi_{(s,u)}(a_{(s,s)},a,z_0) = $
$$
\scriptscriptstyle
\begin{cases}
    \bindist{z_0}{\pflipsyn{s}{u}}{a-a_{(s,s)}}\\ 
    \quad \quad \quad \text{if $l_{01}+\min(l_{11},f_{11})$ even}\\
    \bindist{z_0-1}{\pflipsyn{s}{u}}{a-a_{(s,s)}-1}\\
    \quad \quad \quad \text{if $l_{01}+\min(l_{11},f_{11})$ odd, $a_{(s,s)}\leq a-1$}\\
    0\\
    \quad \quad \quad \text{if $l_{01}+\min(l_{11},f_{11})$ odd, $a_{(s,s)}=a$}\\
\end{cases}
$$
The derivation of $\kappa_s(a_{(s,s)},a,z_0)$ is the same as the one in Appendix \ref{app:trmatrix}, with the only difference being when $a=0$: since this case is not possible, given that by hypothesis at least one flip happens in parity checks that were satisfied before the first iteration, we set $\kappa_s(a_{(s,s)},0,z_0) = 0$. The last modification is in $\Pr\left(\upcfirstit{j} = v-a | j \in \left(\jset{0}{1} \cap \text{supp}(H_{i,:})\right)\right)$, where $\text{supp}(H_{i,:})$ is the set of positions included in the selected parity check $i$. Since these bits are included in a p.c. where $\mathcal{F}=f$, the probability distribution of their upc is the following:
$$
\Pr\left(\upcfirstit{j} = x | j \in \left(\jset{0}{1} \cap \text{supp}(H_{i,:})\right)\right) =
$$
$$
=
\begin{cases}
   \frac{\bindist{v-1}{\punsatz}{\mathtt{u}_1}}{\sum_{x=\thfirstit}^{v-1} \bindist{v-1}{\punsatz}{x}} & \text{if $\thfirstit \leq \mathtt{u}_1 \leq v-1$} \\
   0 & \text{otherwise}
\end{cases}
$$
Once these modifications have been applied, all the transition probabilities for the step $(l_{01},l_{11}) \to (l_{01}+1,l_{11})$ are correctly defined when $l_{01} < f_{01}$.

For $l_{01} \geq f_{01}$, the selected parity check is \textit{not} flipped by hypothesis. Therefore, the definitions of $\chi_{(s,s)}(a_{(s,s)},z_0)$ and $\chi_{(s,u)}(a_{(s,s)},a,z_0)$ become 
$\chi_{(s,u)}(a_{(s,s)},z_0) = $
$$
\scriptscriptstyle
\begin{cases}
    \bindist{r-y-z_0-1}{\pflipsyn{s}{s}}{a_{(s,s)}}\\ 
    \quad \quad \quad \text{if $f_{01}+\min(l_{11},f_{11})$ even}\\
    \bindist{r-y-z_0}{\pflipsyn{s}{s}}{a_{(s,s)}}\\
    \quad \quad \quad \text{if $f_{01}+\min(l_{11},f_{11})$ odd}\\
\end{cases}
$$
and $\chi_{(s,u)}(a_{(s,s)},a,z_0) = $
$$
\scriptscriptstyle
\begin{cases}
    \bindist{z_0}{\pflipsyn{s}{u}}{a-a_{(s,s)}}\\ 
    \quad \quad \quad \text{if $f_{01}+\min(l_{11},f_{11})$ even}\\
    \bindist{z_0-1}{\pflipsyn{s}{u}}{a-a_{(s,s)}}\\
    \quad \quad \quad \text{if $f_{01}+\min(l_{11},f_{11})$ odd}\\
\end{cases}
$$
Moreover, the upc distribution of the bits in such case is the one derived in Appendix \ref{app:trmatrix}, since the selected position is not assumed as included in the (satisfied) parity check. With such modifications, all the transition probabilities for the steps $(l_{01},l_{11}) \to (l_{01}+1,l_{11})$ are correctly defined.

Regarding the steps $(l_{01},l_{11}) \to (l_{01},l_{11}+1)$ where $l_{11} < f_{11}$, the definition of $\chi_{(s,s)}(a_{(s,s)},z_0)$ and $\chi_{(s,u)}(a_{(s,s)},a,z_0)$ is almost the same as in the case of $(l_{01},l_{11}) \to (l_{01}+1,l_{11})$ with $l_{01} < f_{01}$, with the only difference being the condition over which the selected parity check is currently satisfied or not: $\chi_{(s,s)}(a_{(s,s)},z_0) = $
\vspace{2cm}
$$
\scriptscriptstyle
\begin{cases}
    \bindist{r-y-z_0-1}{\pflipsyn{s}{s}}{a_{(s,s)}-1}\\ 
    \quad \quad \quad \text{if $\min(l_{01},f_{01}) + l_{11}$ even, $a_{(s,s)}\geq 1$}\\
    0\\
    \quad \quad \quad \text{if $\min(l_{01},f_{01}) + l_{11}$ even, $a_{(s,s)}=0$}\\
    \bindist{r-y-z_0}{\pflipsyn{s}{s}}{a_{(s,s)}}\\
    \quad \quad \quad \text{if $\min(l_{01},f_{01}) + l_{11}$ odd}\\
\end{cases}
$$
and $\chi_{(s,u)}(a_{(s,s)},a,z_0) = $
$$
\scriptscriptstyle
\begin{cases}
    \bindist{z_0}{\pflipsyn{s}{u}}{a-a_{(s,s)}}\\ 
    \quad \quad \quad \text{if $\min(l_{01},f_{01}) + l_{11}$ even}\\
    \bindist{z_0-1}{\pflipsyn{s}{u}}{a-a_{(s,s)}-1}\\
    \quad \quad \quad \text{if $\min(l_{01},f_{01}) + l_{11}$ odd, $a_{(s,s)}\leq a-1$}\\
    0\\
    \quad \quad \quad \text{if $\min(l_{01},f_{01}) + l_{11}$ odd, $a_{(s,s)}=a$}\\
\end{cases}
$$

The distribution of $\upcfirstit{j}$ for positions $j \in (\jset{1}{1} \cap \text{supp}(H_{i,:}))$, given that the parity check $i$ is satisfied, is:
$$
\Pr\left(\upcfirstit{j} = x | j \in \left(\jset{1}{1} \cap \text{supp}(H_{i,:})\right)\right) =
$$
$$
=
\begin{cases}
   \frac{\bindist{v-1}{\punsato}{\mathtt{u}_1}}{\sum_{x=0}^{\thfirstit-1} \bindist{v-1}{\punsato}{x}} & \text{if $\mathtt{u}_1 \leq \thfirstit-1$} \\
   0 & \text{otherwise}
\end{cases}
$$
When $l_{11} \geq f_{11}$, the definition of $\chi_{(s,s)}(a_{(s,s)},z_0)$ and $\chi_{(s,u)}(a_{(s,s)},a,z_0)$ is almost the same as in the case of $(l_{01},l_{11}) \to (l_{01}+1,l_{11})$ with $l_{01} \geq f_{01}$, with the only difference being the condition
over which the selected parity check is currently satisfied or
not:
$\chi_{(s,u)}(a_{(s,s)},z_0) = $
$$
\scriptscriptstyle
\begin{cases}
    \bindist{r-y-z_0-1}{\pflipsyn{s}{s}}{a_{(s,s)}}\\ 
    \quad \quad \quad \text{if $\min(l_{01},f_{01})+f_{11}$ even}\\
    \bindist{r-y-z_0}{\pflipsyn{s}{s}}{a_{(s,s)}}\\
    \quad \quad \quad \text{if $\min(l_{01},f_{01})+f_{11}$ odd}\\
\end{cases}
$$
and $\chi_{(s,u)}(a_{(s,s)},a,z_0) = $
$$
\scriptscriptstyle
\begin{cases}
    \bindist{z_0}{\pflipsyn{s}{u}}{a-a_{(s,s)}}\\ 
    \quad \quad \quad \text{if $\min(l_{01},f_{01})+f_{11}$ even}\\
    \bindist{z_0-1}{\pflipsyn{s}{u}}{a-a_{(s,s)}}\\
    \quad \quad \quad \text{if $\min(l_{01},f_{01})+f_{11}$ odd}\\
\end{cases}
$$
The distribution of $\upcfirstit{j}$, with $j \in \jset{1}{1}$, is again the one derived in Appendix \ref{app:trmatrix}.

We now move to the case where $f$ is odd. The difference from the analysis performed up to this point is that the selected parity check is now assumed to be unsatisfied before the first iteration. The same assumption can be made regarding the steps where the flips on the selected p.c. happen, meaning that all the first $f_{01}$ steps of the type $(l_{01}, l_{11}) \to (l_{01} + 1, l_{11})$
and all the first $f_{11}$ of the type $(l_{01}, l_{11}) \to (l_{01}, l_{11}+1)$ flip
the selected p.c., while the successive ones do not impact the
parity of the selected equation.

We begin by analyzing the steps $(l_{01},l_{11}) \to (l_{01}+1,l_{11})$ where $l_{01} < f_{01}$. The first difference with respect to the calculations in Appendix \ref{app:trmatrix} is in the derivation of $\chi_{(u,s)}(a_{(u,s)},z_1)$ and $\chi_{(u,u)}(a_{(u,s)},a,z_1)$, denoting the probability that $a_{(u,s)}$ flips happen on parity checks that were unsatisfied before the first iteration and are currently satisfied and that $v - a - a_{(u,s)}$ flips happen on parity checks that were satisfied before the first iteration and are currently unsatisfied (for a fixed value $a$). By hypothesis, we assumed that one of the p.c.s unsatisfied before the first iteration is flipped during the current step (since $l_{01} < f_{01}$). Moreover, said parity check is currently satisfied if $l_{01}+\min(l_{11},f_{11})$ is even, and unsatisfied otherwise. The definition of $\chi_{(u,s)}(a_{(u,s)},z_1)$ and $\chi_{(u,u)}(a_{(u,s)},a,z_1)$, with the inclusion of these information, is
$\chi_{(u,s)}(a_{(u,s)},z_0) = $
$$
\scriptscriptstyle
\begin{cases}
    \bindist{y-z_1-1}{\pflipsyn{u}{s}}{a_{(u,s)}-1}\\ 
    \quad \quad \quad \text{if $l_{01}+\min(l_{11},f_{11})$ even, $a_{(u,s)}\geq 1$}\\
    0\\
    \quad \quad \quad \text{if $l_{01}+\min(l_{11},f_{11})$ even, $a_{(u,s)}=0$}\\
    \bindist{y-z_1}{\pflipsyn{u}{s}}{a_{(u,s)}}\\
    \quad \quad \quad \text{if $l_{01}+\min(l_{11},f_{11})$ odd}\\
\end{cases}
$$
and $\chi_{(u,u)}(a_{(u,s)},a,z_1) = $
$$
\scriptscriptstyle
\begin{cases}
    \bindist{z_1}{\pflipsyn{u}{u}}{v-a-a_{(u,s)}}\\ 
    \quad \quad \quad \text{if $l_{01}+\min(l_{11},f_{11})$ even}\\
    \bindist{z_1-1}{\pflipsyn{u}{u}}{v-a-a_{(u,s)}-1}\\
    \quad \quad \quad \text{if $l_{01}+\min(l_{11},f_{11})$ odd, $a_{(u,s)}\leq v-a-1$}\\
    0\\
    \quad \quad \quad \text{if $l_{01}+\min(l_{11},f_{11})$ odd, $a_{(s,s)}=a$}\\
\end{cases}
$$
The derivation of $\kappa_u(a_{(u,s)},a,z_1)$ is the same as the one in Appendix \ref{app:trmatrix}, with the only difference being when $a=v$: since this case is not possible, given that by hypothesis at least one flip happens in parity checks that were unsatisfied before the first iteration, we set $\kappa_u(a_{(u,s)},v,z_1) = 0$. The last modification is in $\Pr\left(\upcfirstit{j} = v-a | j \in \left(\jset{0}{1} \cap \text{supp}(H_{i,:})\right)\right)$, where $\text{supp}(H_{i,:})$ is the set of positions included in the selected parity check $i$. Since these bits are included in a p.c. where $\mathcal{F}=f$, the probability distribution of their upc is the following:
$$
\Pr\left(\upcfirstit{j} = x | j \in \left(\jset{0}{1} \cap \text{supp}(H_{i,:})\right)\right) =
$$
$$
=
\begin{cases}
   \frac{\bindist{v-1}{\punsatz}{\mathtt{u}_1-1}}{\sum_{x=\thfirstit-1}^{v-1} \bindist{v-1}{\punsatz}{x}} & \text{if $\mathtt{u}_1 \geq \thfirstit$} \\
   0 & \text{otherwise}
\end{cases}
$$
Once these modifications have been applied, all the transition probabilities for the step $(l_{01},l_{11}) \to (l_{01}+1,l_{11})$ are correctly defined when $l_{01} < f_{01}$.

For $l_{01} \geq f_{01}$, the selected parity check is \textit{not} flipped by hypothesis. Therefore, the definitions of $\chi_{(u,s)}(a_{(u,s)},z_1)$ and $\chi_{(u,u)}(a_{(u,s)},a,z_1)$ become 
$\chi_{(u,s)}(a_{(u,s)},z_1) = $
$$
\scriptscriptstyle
\begin{cases}
    \bindist{y-z_1-1}{\pflipsyn{u}{s}}{a_{(u,s)}}\\ 
    \quad \quad \quad \text{if $f_{01}+\min(l_{11},f_{11})$ even}\\
    \bindist{y-z_1}{\pflipsyn{u}{s}}{a_{(u,s)}}\\
    \quad \quad \quad \text{if $f_{01}+\min(l_{11},f_{11})$ odd}\\
\end{cases}
$$
and $\chi_{(u,u)}(a_{(u,s)},a,z_1) = $
$$
\scriptscriptstyle
\begin{cases}
    \bindist{z_1}{\pflipsyn{u}{u}}{v-a-a_{(u,s)}}\\ 
    \quad \quad \quad \text{if $f_{01}+\min(l_{11},f_{11})$ even}\\
    \bindist{z_1-1}{\pflipsyn{u}{u}}{v-a-a_{(u,s)}}\\
    \quad \quad \quad \text{if $f_{01}+\min(l_{11},f_{11})$ odd}\\
\end{cases}
$$
Moreover, the upc distribution of the bits in such case is the one derived in Appendix \ref{app:trmatrix}, since the selected position is not assumed as included in the (unsatisfied) parity check. With such modifications, all the transition probabilities for the steps $(l_{01},l_{11}) \to (l_{01}+1,l_{11})$ are correctly defined.

Regarding the steps $(l_{01},l_{11}) \to (l_{01},l_{11}+1)$ where $l_{11} < f_{11}$, the definition of $\chi_{(u,s)}(a_{(u,s)},z_1)$ and $\chi_{(u,u)}(a_{(u,s)},a,z_1)$ is almost the same as in the case of $(l_{01},l_{11}) \to (l_{01}+1,l_{11})$ with $l_{01} < f_{01}$, with the only difference being the condition over which the selected parity check is currently satisfied or not: 
$\chi_{(u,s)}(a_{(u,s)},z_0) = $
$$
\scriptscriptstyle
\begin{cases}
    \bindist{y-z_1-1}{\pflipsyn{u}{s}}{a_{(u,s)}-1}\\ 
    \quad \quad \quad \text{if $\min(l_{01},f_{01}) + l_{11}$ even, $a_{(u,s)}\geq 1$}\\
    0\\
    \quad \quad \quad \text{if $\min(l_{01},f_{01}) + l_{11}$ even, $a_{(u,s)}=0$}\\
    \bindist{y-z_1}{\pflipsyn{u}{s}}{a_{(u,s)}}\\
    \quad \quad \quad \text{if $\min(l_{01},f_{01}) + l_{11}$ odd}\\
\end{cases}
$$
and $\chi_{(u,u)}(a_{(u,s)},a,z_1) = $
$$
\scriptscriptstyle
\begin{cases}
    \bindist{z_1}{\pflipsyn{u}{u}}{v-a-a_{(u,s)}}\\ 
    \quad \quad \quad \text{if $\min(l_{01},f_{01}) + l_{11}$ even}\\
    \bindist{z_1-1}{\pflipsyn{u}{u}}{v-a-a_{(u,s)}-1}\\
    \quad \quad \quad \text{if $\min(l_{01},f_{01}) + l_{11}$ odd, $a_{(u,s)}\leq v-a-1$}\\
    0\\
    \quad \quad \quad \text{if $\min(l_{01},f_{01}) + l_{11}$ odd, $a_{(s,s)}=a$}\\
\end{cases}
$$
The distribution of $\upcfirstit{j}$ for positions $j \in (\jset{1}{1} \cap \text{supp}(H_{i,:}))$, given that the parity check $i$ is unsatisfied, is:
$$
\Pr\left(\upcfirstit{j} = x | j \in \left(\jset{1}{1} \cap \text{supp}(H_{i,:})\right)\right) =
$$
$$
=
\begin{cases}
   \frac{\bindist{v-1}{\punsato}{\mathtt{u}_1-1}}{\sum_{x=0}^{\thfirstit-2} \bindist{v-1}{\punsato}{x}} & \text{if $1 \leq \mathtt{u}_1 \leq \thfirstit-1$} \\
   0 & \text{otherwise}
\end{cases}
$$
When $l_{11} \geq f_{11}$, the definition of $\chi_{(u,s)}(a_{(u,s)},z_1)$ and $\chi_{(u,u)}(a_{(u,s)},a,z_1)$ is almost the same as in the case of $(l_{01},l_{11}) \to (l_{01}+1,l_{11})$ with $l_{01} \geq f_{01}$, with the only difference being the condition
over which the selected parity check is currently satisfied or
not:
$\chi_{(u,s)}(a_{(u,s)},z_1) = $
$$
\scriptscriptstyle
\begin{cases}
    \bindist{y-z_1-1}{\pflipsyn{u}{s}}{a_{(u,s)}}\\ 
    \quad \quad \quad \text{if $\min(l_{01},f_{01})+f_{11}$ even}\\
    \bindist{y-z_1}{\pflipsyn{u}{s}}{a_{(u,s)}}\\
    \quad \quad \quad \text{if $\min(l_{01},f_{01})+f_{11}$ odd}\\
\end{cases}
$$
and $\chi_{(u,u)}(a_{(u,s)},a,z_1) = $
$$
\scriptscriptstyle
\begin{cases}
    \bindist{z_1}{\pflipsyn{u}{u}}{v-a-a_{(u,s)}}\\ 
    \quad \quad \quad \text{if $\min(l_{01},f_{01})+f_{11}$ even}\\
    \bindist{z_1-1}{\pflipsyn{u}{u}}{v-a-a_{(u,s)}}\\
    \quad \quad \quad \text{if $\min(l_{01},f_{01})+f_{11}$ odd}\\
\end{cases}
$$
The distribution of $\upcfirstit{j}$, with $j \in \jset{1}{1}$, is again the one derived in Appendix \ref{app:trmatrix}.

Upon deriving the definition of the transition probabilities for our Markov process, considering both the cases where $f$ is even or odd, we can compute the p.m.f. of $(\mathcal{Z}_0,\mathcal{Z}_1)_{(\epsilon_{01},\epsilon_{11})}$. Indeed, the desired probability $\Pr(z_0,z_1|f,\epsilon_{01},\epsilon_{11},f_{01},f_{11})$ can be computed as $\Pr\left( (\mathcal{Z}_0=z_0,\mathcal{Z}_1=z_1)_{(\epsilon_{01},\epsilon_{11})} \right)$, applying the presented modifications to the Markov process defined in Appendix \ref{app:trmatrix}.

%% file: appendix3.tex
In Section \ref{sec:model}, the decoding failure rate is estimated by means of $\pflipzz$, $\pflipzo$, $\pflipoz$ and $\pflipoo$, representing the probability of flipping bits in $\jset{0}{0}$, $\jset{0}{1}$, $\jset{1}{0}$ and $\jset{1}{1}$ respectively:
$$
\mathtt{DFR}(y,\epsilon_{01},\epsilon_{11},z_0,z_1) = 
1 -
(1 - \pflipzz)^{n-t-\epsilon_{01}} \cdot 
$$
$$
\cdot 
(\pflipzo)^{\epsilon_{01}}
\cdot
(1-\pflipoz)^{t-\epsilon_{11}}
\cdot
(\pflipoo)^{\epsilon_{11}}
$$
This estimation, in line with the assumption made in  Section \ref{sec:background}, assumes each flipping decision to be taken independently. In the following, we (partly) remove this assumption by taking into account the regularity of the parity check matrix. To this end, we point out that the sum of all the unsatisfied parity check counts is equal to $w \cdot (z_0 + z_1)$. This is true because, for every asserted bit in the syndrome, the $w$ upcs in the corresponding parity check are incremented by one. Additionally, we have that the sum of the unsatisfied parity check counts of incorrect bits after the first iteration (i.e. bits in $\jset{0}{1}$ or $\jset{1}{1}$) is greater than or equal to $z_0+z_1$. This is true because each unsatisfied check contains an odd number of incorrect bits, therefore for each unsatisfied parity check the upc of \textit{at least} one incorrect bit is incremented by one. Refining this line of reasoning, after the first iteration the sum of unsatisfied parity checks that were satisfied before the first iteration is $w \cdot z_0$ across all positions and at least $z_0$ across incorrect positions. Likewise, the sum of unsatisfied parity checks that were unsatisfied before the first iteration is $w \cdot z_1$ across all positions and at least $z_1$ across incorrect positions.

We define $\mu_{00}(x_s,x_u)$ as the probability that a bit $i \in \jset{0}{0}$ appears in $x_s$ unsatisfied parity checks that were satisfied before the first iteration, and $x_u$ unsatisfied parity checks that were unsatisfied before the first iteration, noting that $\upcsecondit{i} = x_s + x_u$. Such function can be computed starting from $\pcheck{00}{s}{u}$ and $\pcheck{00}{u}{u}$, the probability that satisfied and unsatisfied checks become unsatisfied after the first iteration, given that one of the bits included in the check is in $\jset{0}{0}$:
$$
\mu_{00}(x_s,x_u) = \sum_{a=x_u}^{\min(v-x_s, \thfirstit-1)} 
\condprob{\upcfirstit{i}=a}{i \in \jset{0}{0}} \cdot 
$$
$$
\cdot \bindist{v-a}{\pcheck{00}{s}{u}}{x_s}
\cdot \bindist{  a}{\pcheck{00}{u}{u}}{x_u}
$$
The derivation of $\mu_{01}(x_s,x_u)$, $\mu_{10}(x_s,x_u)$ and $\mu_{11}(x_s,x_u)$ is equivalent up to a matter of indexes. For $i \in \jset{0}{1}$:
$$
\mu_{01}(x_s,x_u) = \sum_{a=\max(x_u, \thfirstit)}^{\min(v-x_s)} 
\condprob{\upcfirstit{i}=a}{i \in \jset{0}{1}} \cdot 
$$
$$
\cdot \bindist{v-a}{\pcheck{01}{s}{u}}{x_s}
\cdot \bindist{  a}{\pcheck{01}{u}{u}}{x_u}
$$
For $i \in \jset{1}{0}$:
$$
\mu_{10}(x_s,x_u) = \sum_{a=\max(x_u, \thfirstit)}^{\min(v-x_s)} 
\condprob{\upcfirstit{i}=a}{i \in \jset{1}{0}} \cdot 
$$
$$
\cdot \bindist{v-a}{\pcheck{10}{s}{u}}{x_s}
\cdot \bindist{  a}{\pcheck{10}{u}{u}}{x_u}
$$
For $i \in \jset{1}{1}$:
$$
\mu_{11}(x_s,x_u) = \sum_{a=x_u}^{\min(v-x_s, \thfirstit-1)} 
\condprob{\upcfirstit{i}=a}{i \in \jset{1}{1}} \cdot 
$$
$$
\cdot \bindist{v-a}{\pcheck{11}{s}{u}}{x_s}
\cdot \bindist{  a}{\pcheck{11}{u}{u}}{x_u}
$$
In the following, we denote as $\Omega^{(s)}_{(l_{00},l_{10})}$ the r.v. associated to the sum of the unsatisfied parity check counts, including only parity equations that were satisfied before the first iteration, computed over a subset of the $n$ positions including $l_{00}$ bits in $\jset{0}{0}$ and $l_{10}$ bits in $\jset{1}{0}$. Likewise, $\Omega^{(u)}_{(l_{00},l_{10})}$ is the r.v. associated to the sum of the unsatisfied parity check counts, including only parity equations that were \textit{unsatisfied} before the first iteration, computed over a subset of the $n$ positions including $l_{00}$ bits in $\jset{0}{0}$ and $l_{10}$ bits in $\jset{1}{0}$. Moreover, we define as $\Theta^{(s)}_{(l_{01},l_{11})}$ the r.v. associated to the sum of the unsatisfied parity check counts, including only parity equations that were satisfied before the first iteration, computed over a subset of the $n$ positions including $l_{01}$ bits in $\jset{0}{1}$ and $l_{11}$ bits in $\jset{1}{1}$. Likewise, $\Theta^{(u)}_{(l_{01},l_{11})}$ is the r.v. associated to the sum of the unsatisfied parity check counts, including only parity equations that were \textit{unsatisfied} before the first iteration, computed over a subset of the $n$ positions including $l_{01}$ bits in $\jset{0}{1}$ and $l_{11}$ bits in $\jset{1}{1}$.
The two random variables $\left(\Omega^{(s)}_{(l_{00},l_{10})},\Theta^{(s)}_{(l_{01},l_{11})}\right)$ are not completely independent: due to the regularity of the parity check matrix $H$, we have that $\Omega^{(s)}_{(\epsilon_{00},\epsilon_{10})}+\Theta^{(s)}_{(\epsilon_{01},\epsilon_{11})} = w \cdot z_0$, with $\Theta^{(s)}_{(\epsilon_{01},\epsilon_{11})} \geq z_0$ (where $\epsilon_{00} = |\jset{0}{0}|$ and $\epsilon_{10} = |\jset{1}{0}|$). For the same reasons, we have $\Omega^{(u)}_{(\epsilon_{00},\epsilon_{10})}+\Theta^{(u)}_{(\epsilon_{01},\epsilon_{11})} = w \cdot z_1$, with $\Theta^{(u)}_{(\epsilon_{01},\epsilon_{11})} \geq z_1$. Finally, $\mathtt{E}^{(\texttt{f})}_{(l_{00},l_{10})}$ is the event of a decoding failure given by an erroneous flip in subsets of $\jset{0}{0}$ and $\jset{1}{0}$ of size $l_{00}$ and $l_{10}$; $\mathtt{E}^{(\texttt{m})}_{(l_{01},l_{11})}$ is the event of a decoding failure given by erroneously maintaining bits in subsets of $\jset{0}{1}$ and $\jset{1}{1}$ of size $l_{01}$ and $l_{11}$ ; $\sevent{\texttt{fail}}$ is the event of a decoding failure given by any incorrect decision taken among all $n$ bits.
In the following, we derive the joint p.m.f. of $\left(\Omega^{(s)}_{(\epsilon_{00},\epsilon_{10})},\Omega^{(u)}_{(\epsilon_{00},\epsilon_{10})},\mathtt{E}^{(\texttt{f})}_{(\epsilon_{00},\epsilon_{10})}\right)$, 
$\left(\Theta^{(s)}_{(\epsilon_{01},\epsilon_{11})},\Theta^{(u)}_{(\epsilon_{01},\epsilon_{11})},\mathtt{E}^{(\texttt{m})}_{(\epsilon_{01},\epsilon_{11})}\right)$, and the probability $\Pr(\sevent{\texttt{fail}})$ introducing the regularity constraint.

The p.m.f. of $\left(\Omega^{(s)}_{(\epsilon_{00},\epsilon_{10})},\Omega^{(u)}_{(\epsilon_{00},\epsilon_{10})},\mathtt{E}^{(\texttt{f})}_{(\epsilon_{00},\epsilon_{10})}\right)$ can be computed by means of a Markov process. The initial state considers the two subsets of $\jset{0}{0}$ and $\jset{1}{0}$ to be empty: the sums computed over such subsets is equal to $0$, and no incorrect flip can happen in such subsets:
$$
\Pr\left(
\Omega^{(s)}_{(0,0)}=0, \Omega^{(u)}_{(0,0)}=0, \neg\mathtt{E}^{(\texttt{f})}_{(0,0)}
\right) = 1
$$
One step of the Markov process corresponds to either the addition of one bit in $\jset{0}{0}$ to the corresponding subset (step $(l_{00},l_{10}) \to (l_{00}+1,l_{10})$), or the addition of one bit in $\jset{1}{0}$ to the corresponding subset (step $(l_{00},l_{10}) \to (l_{00},l_{10}+1)$). 

We begin by modeling the transition probabilities for the step $(l_{00},l_{10}) \to (l_{00}+1,l_{10})$. Given a position $i \in \jset{0}{0}$, the probability that $\Omega^{(s)}$ is increased by $x_s$ (meaning $\Omega^{(s)}_{(l_{00}+1,l_{10})} = \Omega^{(s)}_{(l_{00},l_{10})} + x_s$) and $\Omega^{(u)}$ is increased by $x_u$ (meaning $\Omega^{(u)}_{(l_{00}+1,l_{10})} = \Omega^{(u)}_{(l_{00},l_{10})} + x_u$) is $\mu_{00}(x_s,x_u)$. Additionally, the event of having an incorrect flip among the $l_{00} + 1 +l_{10}$ selected bits ($\mathtt{E}^{(\texttt{f})}_{(l_{00}+1,l_{10})}$) is bound to the fact that either there has already been an erroneous flip in the previously selected $l_{00} + l_{10}$ positions ($\mathtt{E}^{(\texttt{f})}_{(l_{00},l_{10})}$), or that the upc of the currently selected bit is greater that or equal to the threshold $\thsecondit$ ($x_s + x_u \geq \thsecondit$). The event $\neg\mathtt{E}^{(\texttt{f})}_{(l_{00}+1,l_{10})}$ requires instead that both the previously stated conditions are false. We can therefore derive the transition probabilities, depending on the truthness of $\mathtt{E}^{(\texttt{f})}_{(l_{00}+1,l_{10})}$ and $\mathtt{E}^{(\texttt{f})}_{(l_{00},l_{10})}$. If $\mathtt{E}^{(\texttt{f})}_{(l_{00}+1,l_{10})}$ and $\mathtt{E}^{(\texttt{f})}_{(l_{00},l_{10})}$ are both false:
$$
\scriptstyle
\Pr\left(
\Omega^{(s)}_{(l_{00}+1,l_{10})}=a_s+x_s, \Omega^{(u)}_{(l_{00}+1,l_{10})}=a_u+x_u, \neg\mathtt{E}^{(\texttt{f})}_{(l_{00}+1,l_{10})}
\right|
$$
$$ 
\scriptstyle
\left| \Omega^{(s)}_{(l_{00},l_{10})}=a_s, \Omega^{(u)}_{(l_{00},l_{10})}=a_u, \neg\mathtt{E}^{(\texttt{f})}_{(l_{00},l_{10})}
\right)
= 
$$
$$
\scriptstyle
=
\begin{cases}
    \mu_{00}(x_s,x_u) & \text{if $x_s+x_u < \thsecondit$}\\
    0 & \text{otherwise}
\end{cases}
$$
If $\mathtt{E}^{(\texttt{f})}_{(l_{00}+1,l_{10})}$ is false and $\mathtt{E}^{(\texttt{f})}_{(l_{00},l_{10})}$ is true, then the transition is not possible since an erroneous flip has already happened:
$$
\scriptstyle
\Pr\left(
\Omega^{(s)}_{(l_{00}+1,l_{10})}=a_s+x_s, \Omega^{(u)}_{(l_{00}+1,l_{10})}=a_u+x_u, \neg\mathtt{E}^{(\texttt{f})}_{(l_{00}+1,l_{10})}
\right|
$$
$$ 
\scriptstyle
\left| \Omega^{(s)}_{(l_{00},l_{10})}=a_s, \Omega^{(u)}_{(l_{00},l_{10})}=a_u, \mathtt{E}^{(\texttt{f})}_{(l_{00},l_{10})}
\right)
= 0
$$
If $\mathtt{E}^{(\texttt{f})}_{(l_{00}+1,l_{10})}$ is true and $\mathtt{E}^{(\texttt{f})}_{(l_{00},l_{10})}$ is false:
$$
\scriptstyle
\Pr\left(
\Omega^{(s)}_{(l_{00}+1,l_{10})}=a_s+x_s, \Omega^{(u)}_{(l_{00}+1,l_{10})}=a_u+x_u, \mathtt{E}^{(\texttt{f})}_{(l_{00}+1,l_{10})}
\right|
$$
$$ 
\scriptstyle
\left| \Omega^{(s)}_{(l_{00},l_{10})}=a_s, \Omega^{(u)}_{(l_{00},l_{10})}=a_u, \neg\mathtt{E}^{(\texttt{f})}_{(l_{00},l_{10})}
\right)
= 
$$
$$
\scriptstyle
=
\begin{cases}
    \mu_{00}(x_s,x_u) & \text{if $x_s+x_u \geq \thsecondit$}\\
    0 & \text{otherwise}
\end{cases}
$$
Finally, if $\mathtt{E}^{(\texttt{f})}_{(l_{00}+1,l_{10})}$ is and $\mathtt{E}^{(\texttt{f})}_{(l_{00},l_{10})}$ are both true:
$$
\scriptstyle
\Pr\left(
\Omega^{(s)}_{(l_{00}+1,l_{10})}=a_s+x_s, \Omega^{(u)}_{(l_{00}+1,l_{10})}=a_u+x_u, \mathtt{E}^{(\texttt{f})}_{(l_{00}+1,l_{10})}
\right|
$$
$$ 
\scriptstyle
\left| \Omega^{(s)}_{(l_{00},l_{10})}=a_s, \Omega^{(u)}_{(l_{00},l_{10})}=a_u, \mathtt{E}^{(\texttt{f})}_{(l_{00},l_{10})}
\right)
= \mu_{00}(x_s,x_u)
$$
These formula describe the transition probabilities modeling the addition of a bit $i \in \jset{0}{0}$ to the corresponding subset (step $(l_{00},l_{10}) \to (l_{00}+1,l_{10})$). The transition probabilities for the step $(l_{00},l_{10}) \to (l_{00},l_{10}+1)$ are exactly the same, with the only difference being $\mu_{10}(x_s,x_u)$ instead of $\mu_{00}(x_s,x_u)$.

The initial state and the transition probabilities fully define the Markov process, in turn allowing us to derive the joint p.m.f. of $\left(\Omega^{(s)}_{(\epsilon_{00},\epsilon_{10})},\Omega^{(u)}_{(\epsilon_{00},\epsilon_{10})},\mathtt{E}^{(\texttt{f})}_{(\epsilon_{00},\epsilon_{10})}\right)$.

The joint p.m.f. of $\left(\Theta^{(s)}_{(\epsilon_{01},\epsilon_{11})},\Theta^{(u)}_{(\epsilon_{01},\epsilon_{11})},\mathtt{E}^{(\texttt{m})}_{(\epsilon_{01},\epsilon_{11})}\right)$ can be computed through an analogous line of reasoning. We define yet another Markov process where the initial state considers the two subsets of $\jset{0}{1}$ and $\jset{1}{1}$ to be empty. The upc sums computed over such subsets is equal to $0$, and no missing flip can happen in such subsets:
$$
\Pr\left(
\Theta^{(s)}_{(0,0)}=0, \Theta^{(u)}_{(0,0)}=0, \neg\mathtt{E}^{(\texttt{m})}_{(0,0)}
\right) = 1
$$
One step of the Markov process corresponds to either the addition of one bit in $\jset{0}{1}$ to the corresponding subset (step $(l_{01},l_{11}) \to (l_{01}+1,l_{11})$), or the addition of one bit in $\jset{1}{1}$ to the corresponding subset (step $(l_{01},l_{11}) \to (l_{01},l_{11}+1)$). 

We begin by modeling the transition probabilities for the step $(l_{01},l_{11}) \to (l_{01}+1,l_{11})$. Through the same reasoning as for $\left(\Omega^{(s)}_{(l_{00},l_{10})},\Omega^{(u)}_{(l_{00},l_{10})},\mathtt{E}^{(\texttt{f})}_{(l_{00},l_{10})}\right)$, the transition probabilities of $\left(\Theta^{(s)}_{(l_{01},l_{11})},\Theta^{(u)}_{(l_{01},l_{11})},\mathtt{E}^{(\texttt{m})}_{(l_{01},l_{11})}\right)$ depend on the truthness of $\mathtt{E}^{(\texttt{m})}_{(l_{01}+1,l_{11})}$ and $\mathtt{E}^{(\texttt{m})}_{(l_{01},l_{11})}$. If $\mathtt{E}^{(\texttt{m})}_{(l_{01}+1,l_{11})}$ and $\mathtt{E}^{(\texttt{m})}_{(l_{01},l_{11})}$ are both false:
$$
\scriptstyle
\Pr\left(
\Theta^{(s)}_{(l_{01}+1,l_{11})}=a_s+x_s, \Theta^{(u)}_{(l_{01}+1,l_{11})}=a_u+x_u, \neg\mathtt{E}^{(\texttt{m})}_{(l_{01}+1,l_{11})}
\right|
$$
$$ 
\scriptstyle
\left| \Theta^{(s)}_{(l_{01},l_{11})}=a_s, \Theta^{(u)}_{(l_{01},l_{11})}=a_u, \neg\mathtt{E}^{(\texttt{m})}_{(l_{01},l_{11})}
\right)
= 
$$
$$
\scriptstyle
=
\begin{cases}
    \mu_{01}(x_s,x_u) & \text{if $x_s+x_u \geq \thsecondit$}\\
    0 & \text{otherwise}
\end{cases}
$$
\begin{figure*}[!t]
\begin{center}
    \input{figures/tikz_pic_syn_dfr}
\end{center}
    \caption{Numerical validation of the model of the decoding failure rate 
    conditioned on the syndrome weight $y$, employing the following 
    parameter set: $n=2,990$, $\frac{k}{n}=\frac{1}{2}$, $v=23$, $t=30$,
    $\mathtt{th}_{(2)}=\lceil \frac{v+1}{2} \rceil$ fixed. Numerical results 
    obtained with $10^8$ random samples for each threshold choice. 
    Solid and dotted lines are the model, crosses are numerical simulations.}
    \label{fig:syndfr}
\end{figure*}
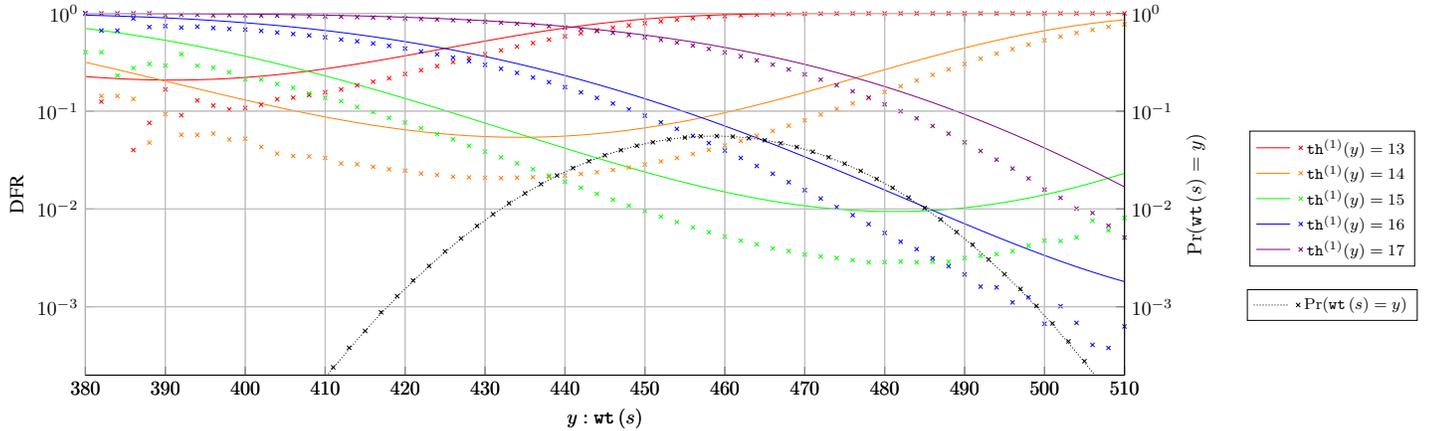
If $\mathtt{E}^{(\texttt{m})}_{(l_{01}+1,l_{11})}$ is false and $\mathtt{E}^{(\texttt{m})}_{(l_{01},l_{11})}$ is true, then the transition is not possible since a missing flip has already happened:
$$
\scriptstyle
\Pr\left(
\Theta^{(s)}_{(l_{01}+1,l_{11})}=a_s+x_s, \Theta^{(u)}_{(l_{01}+1,l_{11})}=a_u+x_u, \neg\mathtt{E}^{(\texttt{m})}_{(l_{01}+1,l_{11})}
\right|
$$
$$ 
\scriptstyle
\left| \Theta^{(s)}_{(l_{01},l_{11})}=a_s, \Theta^{(u)}_{(l_{01},l_{11})}=a_u, \mathtt{E}^{(\texttt{m})}_{(l_{01},l_{11})}
\right)
= 0
$$
If $\mathtt{E}^{(\texttt{m})}_{(l_{01}+1,l_{11})}$ is true and $\mathtt{E}^{(\texttt{m})}_{(l_{01},l_{11})}$ is false:
$$
\scriptstyle
\Pr\left(
\Theta^{(s)}_{(l_{01}+1,l_{11})}=a_s+x_s, \Theta^{(u)}_{(l_{01}+1,l_{11})}=a_u+x_u, \mathtt{E}^{(\texttt{m})}_{(l_{01}+1,l_{11})}
\right|
$$
$$ 
\scriptstyle
\left| \Theta^{(s)}_{(l_{01},l_{11})}=a_s, \Theta^{(u)}_{(l_{01},l_{11})}=a_u, \neg\mathtt{E}^{(\texttt{m})}_{(l_{01},l_{11})}
\right)
= 
$$
$$
\scriptstyle
=
\begin{cases}
    \mu_{01}(x_s,x_u) & \text{if $x_s+x_u < \thsecondit$}\\
    0 & \text{otherwise}
\end{cases}
$$
Finally, if $\mathtt{E}^{(\texttt{m})}_{(l_{01}+1,l_{11})}$ is and $\mathtt{E}^{(\texttt{m})}_{(l_{01},l_{11})}$ are both true:
$$
\scriptstyle
\Pr\left(
\Theta^{(s)}_{(l_{01}+1,l_{11})}=a_s+x_s, \Theta^{(u)}_{(l_{01}+1,l_{11})}=a_u+x_u, \mathtt{E}^{(\texttt{m})}_{(l_{01}+1,l_{11})}
\right|
$$
$$ 
\scriptstyle
\left| \Theta^{(s)}_{(l_{01},l_{11})}=a_s, \Theta^{(u)}_{(l_{01},l_{11})}=a_u, \mathtt{E}^{(\texttt{m})}_{(l_{01},l_{11})}
\right)
= \mu_{01}(x_s,x_u)
$$

These formula describe the transition probabilities modeling the addition of a bit $i \in \jset{0}{1}$ to the corresponding subset (step $(l_{01},l_{11}) \to (l_{01}+1,l_{11})$). The transition probabilities for the step $(l_{01},l_{11}) \to (l_{01},l_{11}+1)$ are exactly the same, with the only difference being $\mu_{11}(x_s,x_u)$ instead of $\mu_{01}(x_s,x_u)$.

The initial state and the transition probabilities fully define the Markov process, in turn allowing us to derive the joint p.m.f. of $\left(\Theta^{(s)}_{(\epsilon_{01},\epsilon_{11})},\Theta^{(u)}_{(\epsilon_{01},\epsilon_{11})},\mathtt{E}^{(\texttt{m})}_{(\epsilon_{01},\epsilon_{11})}\right)$.

We now have all the tools required for computing $\Pr\left(\sevent{\texttt{fail}}\right)$ introducing the regularity constraint. We remind that these constraints impose $\Omega^{(s)}_{(\epsilon_{00},\epsilon_{10})}+\Theta^{(s)}_{(\epsilon_{01},\epsilon_{11})} = w \cdot z_0$,  $\Theta^{(s)}_{(\epsilon_{01},\epsilon_{11})} \geq z_0$, $\Omega^{(u)}_{(\epsilon_{00},\epsilon_{10})}+\Theta^{(u)}_{(\epsilon_{01},\epsilon_{11})} = w \cdot z_1$, and $\Theta^{(u)}_{(\epsilon_{01},\epsilon_{11})} \geq z_1$: we will denote as $\sevent{v,w}$ the event of such constraints taking place.  Additionally, $\sevent{\texttt{fail}} = 
\mathtt{E}^{(\texttt{f})}_{(\epsilon_{00},\epsilon_{10})}
\cup \mathtt{E}^{(\texttt{m})}_{(\epsilon_{01},\epsilon_{11})}$.
We can now derive $\Pr\left(\sevent{\texttt{fail}} | \sevent{v,w} \right)$ as:
$$
\Pr\left(\sevent{\texttt{fail}} | \sevent{v,w} \right) = 
$$
$$
=
\frac{
\Pr\left(\sevent{\texttt{fail}} \cap \sevent{v,w} \right)
}{
\Pr\left(\sevent{\texttt{fail}} \cap \sevent{v,w} \right) + 
\Pr\left(\neg\sevent{\texttt{fail}} \cap \sevent{v,w} \right)
}
$$
We can explicitly compute $\Pr\left(\sevent{\texttt{fail}} \cap \sevent{v,w} \right)$ and 
$\Pr\left(\neg \sevent{\texttt{fail}} \cap \sevent{v,w} \right)$ using the previously derived distributions. The probability of a fail happening along with the regularity constraints is:
$$
\Pr\left(\sevent{\texttt{fail}} \cap \sevent{v,w} \right) = 
\sum_{\theta_s \geq z_1}
\quad
\sum_{\theta_u \geq z_1}
\quad
\sum_{\omega_s | \omega_s + \theta_s = w z_0}
$$
$$
\sum_{\omega_u | \omega_u + \theta_u = w z_1}
\rho_\texttt{f}\left(\omega_s,\omega_u\right) + 
\rho_\texttt{m}\left(\theta_s,\theta_u\right) -
\rho_\texttt{f}\left(\omega_s,\omega_u\right) \cdot
\rho_\texttt{m}\left(\theta_s,\theta_u\right)
$$

The probability of a decoding success along with the regularity constraints is:
$$
\Pr\left(\neg\sevent{\texttt{fail}} \cap \sevent{v,w} \right) = 
\sum_{\theta_s \geq z_1}
\quad
\sum_{\theta_u \geq z_1}
\quad
\sum_{\omega_s | \omega_s + \theta_s = w z_0}
$$
$$
\sum_{\omega_u | \omega_u + \theta_u = w z_1}
(1-\rho_\texttt{f}\left(\omega_s,\omega_u\right))\cdot 
(1-\rho_\texttt{m}\left(\theta_s,\theta_u\right))
$$
Where $\rho_\texttt{f}\left(\omega_s,\omega_u\right)$ denotes the probability of an incorrect flip happening when $\Omega^{(s)}_{(\epsilon_{00},\epsilon_{10})}= \omega_s$ and $\Omega^{(u)}_{(\epsilon_{00},\epsilon_{10})}= \omega_u$, while $\rho_\texttt{m}\left(\theta_s,\theta_u\right)$ is the probability of a missing flip happening when $\Theta^{(s)}_{(\epsilon_{01},\epsilon_{11})}=\theta_s$ 
and $\Theta^{(u)}_{(\epsilon_{01},\epsilon_{11})}=\theta_u$:
$$
\rho_\texttt{f}\left(\omega_s,\omega_u\right) = 
\Pr\left( \Omega^{(s)}_{(\epsilon_{00},\epsilon_{10})}= \omega_s,\Omega^{(u)}_{(\epsilon_{00},\epsilon_{10})}= \omega_u,\mathtt{E}^{(\texttt{f})}_{(\epsilon_{00},\epsilon_{10})}\right)
$$

$$
\rho_\texttt{m}\left(\theta_s,\theta_u\right) = 
\Pr\left(\Theta^{(s)}_{(\epsilon_{01},\epsilon_{11})}=\theta_s,\Theta^{(u)}_{(\epsilon_{01},\epsilon_{11})}=\theta_u,\mathtt{E}^{(\texttt{m})}_{(\epsilon_{01},\epsilon_{11})}\right)
$$
We have now successfully computed the probability $\Pr\left(\sevent{\texttt{fail}} | \sevent{v,w} \right)$, which is an approximation of the decoding failure rate that takes into account the $(v,w)$-regularity of the parity check matrix $H$. We point out that, since the previous steps of the model assume the parity checks to be independent, employing this technique alongside the rest of the model \textit{as is} does not guarantee a conservative estimation of the DFR. This heuristic is therefore not suited for conservative threshold evaluation, but can be fruitfully employed for threshold selection.

%% file: figures/tikz_pic_syn_dfr.tex
\pgfkeys{/pgf/number format/.cd,1000 sep={}}
\begin{tikzpicture}[scale=0.75]
  \begin{axis}[%
               width=20cm, 
               height=8cm,
               legend columns=2,
               xmin = 380,
               xmax = 510,
               grid = major,
               ymode= log,
               log basis y=10,
               ymin = 2e-4,
               ymax = 1,
               ytick = {1,1e-1,1e-2,1e-3,1e-4},
               legend style={at={(1.2,0.3)},anchor=south, font=\footnotesize},
               mark size=1.5pt,
               xlabel={$y:\weight{s}$},
               ylabel={DFR},
               line width=0.5pt,
               cycle list name=waterfall-floor,
               xticklabel style={
                 /pgf/number format/fixed,
                 /pgf/number format/precision=5
               },
scaled x ticks=false]

   \foreach \thchoice in {13,14,15,16,17} {
     \addplot table [x expr = \thisrow{w},
                     y expr= \thisrow{dfr_w},
                     restrict y to domain=-18.1:1, 
                     col sep = comma] {data/syn_dfr/model/syndfr_2_1495_23_30_\thchoice_12.csv};
     \addlegendentryexpanded{};
     \addplot table [x expr = \thisrow{sw},
                     y expr = \thisrow{dfr}, 
                     restrict y to domain=-18.3:1, 
                     col sep = comma] {data/syn_dfr/sim_10e8/syndfr_2_1495_23_30_\thchoice_12.csv};
    \addlegendentryexpanded{$\thfirstit=\thchoice$};
   }

\end{axis}

\begin{axis}[
               width=20cm, 
               height=8cm,
               legend columns=2,
               xmin = 380,
               xmax = 510,
               grid = major,
               ymode= log,
               log basis y=10,
               ymin = 2e-4,
               ymax = 1,
               ytick = {1,1e-1,1e-2,1e-3,1e-4},
               legend style={at={(1.2,0.15)},anchor=south, font=\footnotesize},
               mark size=1.5pt,
               xlabel={$y:\weight{s}$},
               ylabel={$\Pr(\weight{s}=y)$},
               line width=0.5pt,
                axis x line*=bottom,
                axis y line*=right,
scaled x ticks=false]

   \addplot[black, densely dotted] table [x expr = \thisrow{w_end_val},
                     y expr= \thisrow{prob},
                     restrict y to domain=-18.1:1, 
                     col sep = comma] {data/sw1495_model.csv};
     \addlegendentryexpanded{};
     \addplot[black,only marks, mark=x] table [x expr = \thisrow{w_end_val},
                     y expr = \thisrow{prob}, 
                     restrict y to domain=-18.3:1, 
                     col sep = comma] {data/sw1495_sim.csv};
    \addlegendentryexpanded{$\Pr(\weight{s}=y)$};

\end{axis}

\end{tikzpicture}

%% file: appendix4.tex
\begin{figure*}[!t]
\begin{center}
    \subfloat[$v=9$, $t \in \{10,15,20,...,50\}$\label{fig:error_sweep}]{
      \input{figures/tikz_pic_error_sweep}
    }
    \subfloat[$v=9$, $t \in \{10,15,20,...,50\}$\label{fig:error_sweep_comparison}]{
       \input{figures/tikz_pic_error_comparison}
    }
\end{center}
\caption{Two iterations DFR estimated values for $(v,2v)$-regular LDPC codes
  with rate $\frac{k}{n}=\frac{1}{2}$, parallel decoder employing thresholds derived from our model (solid lines) against fixed thresholds with majority voting $\mathtt{th}^{(1)}=\mathtt{th}^{(2)}=\lceil \frac{v+1}{2} \rceil$ (dashed lines), as in \cite{DBLP:conf/isit/AnnechiniBP24}, Figure 2.}
 \vspace{-0.5cm}
 
\end{figure*}
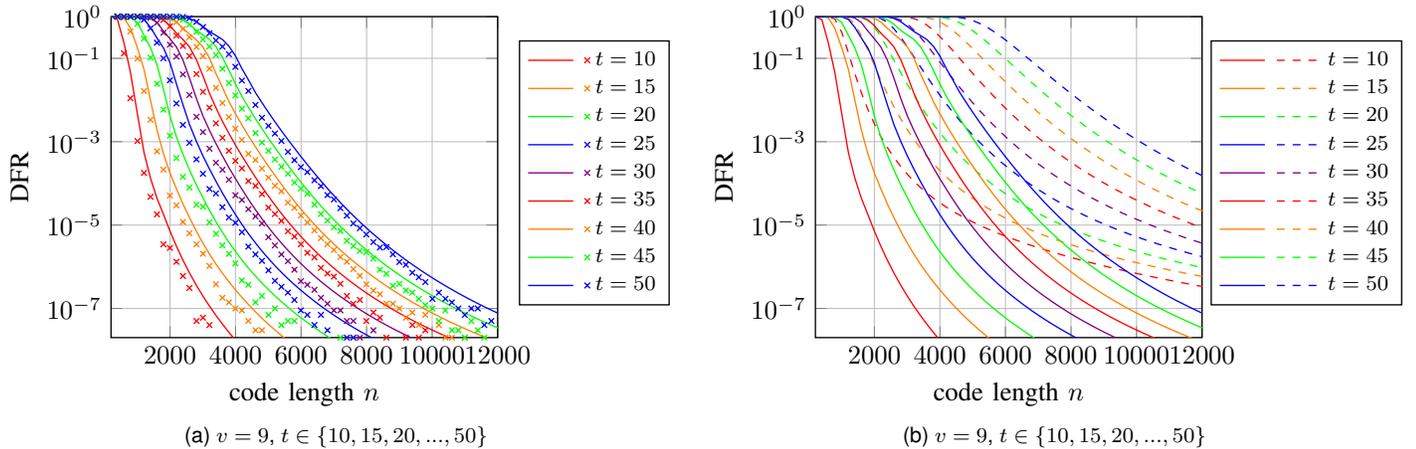

This appendix provides additional numerical results on the effectiveness
of our threshold selection approach.

Figure \ref{fig:syndfr} reports the comparison between our model and numerical 
simulations of the decoding failure rate conditioned on the syndrome weight 
before the first iteration of the decoder. 
The figure shows the failure probability after two iterations, employing 
different fixed threshold choices $\thfirstit$ for every syndrome weight $y$. 
Moreover, the figure displays on the right axis the probability distribution 
(both modeled and simulated) of the syndrome weight $y$ before the first 
iteration. 
The numerical simulations show that, for each threshold value, the decoding 
failure rate follows an inverted bell curve, where syndrome weights that are 
too low or too high cause the decoder to fail more often. 
Our model is able to predict this pattern, up to a small conservative margin. 
Regarding the threshold selection process, the optimal threshold for each 
value of $y$ is the one that minimizes the DFR for the said value of $y$. 
The curve resulting from the optimal threshold selection corresponds to 
the envelope of the various curves, where at each point the threshold that 
minimizes the DFR is selected: we are thus interested in the intersections 
between each curve pair, delimiting the range of $y$ where each 
threshold value is optimal. The intersections between the curves generated 
by our model line up verically almost perfectly with the ones generated 
by numerical simulations.

As a side note, we point out that our model confirms the hypothesis that, 
for the first iteration, the optimal threshold is roughly proportional 
to the syndrome weight: the higher the syndrome weight, the higher the 
optimal threshold. 
This can be seen in Figure \ref{fig:syndfr}, as the line with the lowest 
DFR corresponds to higher values of $\thfirstit$ as $y$ increases, and the
crossver points between curves characterized by fixed thresholds differing
by one unit almost lie on a single line. 

As a further remark, we note that both our model and numerical simulations show
that, whenever the syndrome weight is too low, the decoder is unable to guess 
the error vector \textit{regardless of the threshold choice} at the first 
iteration. We ascribe this fact to the information loss intrinsic to the fact
that the syndrome contains only parity check values, and a very low weight 
syndrome (unless its weight is a multiple of $v$) is likely coming from a 
significant amount of cancellations in parity check equations.
\begin{table}
\caption{Table reporting the average number of discrepancies left over by a 
two iteration parallel decoder employing our criteria for $\mathtt{th}^{(1)}, 
\mathtt{th}^{(2)}$,in comparison with the one from~\cite{BIKE}\label{tab:leftovers_bike}}
 \vspace{-0.4cm}
\begin{center}
\input{figures/tab_BIKE_comparison} 
\end{center}
 \vspace{-0.5cm}
\end{table}
\begin{table}
\caption{Table reporting the improvement in the DFR for a two-iterations
parallel bit flipping decoder on code with parameters from~\cite{LEDA}, 
security category $1$ and different rates.\label{tab:leda}}
 \vspace{-0.4cm}
\begin{center}
  \input{figures/tab_dfr_leda}
 \end{center}
 \vspace{-0.5cm}
\end{table}
Figures~\ref{fig:error_sweep} and~\ref{fig:error_sweep_comparison} report
the results of the numerical validation of the goodness of fit of our DFR 
prediction model under variable threshold choices, and the gain of employing our
threshold selection approach considering rate $\frac{1}{2}$ codes, of lengths
in $\{2000,\ldots,12000\}$, column weight $v=9$ and error weight 
$t\in\{10,15,\ldots,50\}$. As it can be seen, our model provides a
good fit for the actual numerical DFR results, and we retain the three orders of magnitude gain, present in the exploration of different code densities in Figure~\ref{fig:density_sweep_comparison}.

Willing to compare our syndrome weight dependent threshold selection approach
with the one in~\cite{BIKE}, we choose the average number of leftover
discrepancies after the second iteration, i.e., $\weight{\bar{d}^{(2)}}$ as
the figure of merit, as BIKE employs a seven iteration decoder, making a
direct DFR comparison unwieldy.
In addition to the results reported in Figure~\ref{fig:bike_comparison}, 
sweeping on a range of values for $n$ close to the BIKE parameters for 
security category 1, 

Table~\ref{tab:leftovers_bike} reports the average weight of the discrepancy vector
after the second iteration $\weight{\bar{d}^{(2)}}$ as a metric of performance of the decoder. Our threshold selection approach
improves substantially on the choice from~\cite{BIKE}, reducing
the average number of discrepancies by a factor ranging between
$3\cdot 10^2$ and $3.4 \cdot 10^6$.
Finally, Table~\ref{tab:leda} reports the improvement achievable on the 
two-iterations decoding failure rate for the code parameters employed in LEDAcrypt~\cite{LEDA}, security category 1, and different rates.
LEDAcrypt employs natively a two-iterations, fixed threshold parallel decoder.
Our threshold selection approach allows us to gain roughly three decades on
the DFR with respect to the fixed threshold choice of~\cite{extended}.

%% file: figures/tikz_pic_error_sweep.tex
\pgfkeys{/pgf/number format/.cd,1000 sep={}}
\begin{tikzpicture}
  \begin{axis}[scale=0.75,
               legend columns=2,
               xmin = 200,
               xmax = 12000,
               grid = major,
               ymode= log,
               log basis y=10,
               ymin = 2e-8,
               ymax = 1,
               ytick = {1,1e-1,1e-3,1e-5,1e-7},
               legend style={at={(1.25,0.1)},anchor=south, font=\footnotesize},
               mark size=1.5pt,
               xlabel={code length $n$},
               ylabel={DFR},
               line width=0.5pt,
               cycle list name=waterfall-floor,
               xticklabel style={
                 /pgf/number format/fixed,
                 /pgf/number format/precision=5
               },
scaled x ticks=false]
   \foreach \tvalue in {10,15,20,...,50} {
     \addplot table [x expr = \thisrow{r}*2,
                     y expr= \thisrow{dfr},
                     restrict y to domain=-18.1:1, 
                     col sep = comma] {data/error_sweep/model/DFR_n0_2_v_9_t_\tvalue.csv};
     \addlegendentryexpanded{};
     \addplot table [x expr = \thisrow{r}*2,
                     y expr = \thisrow{dfr}, 
                     restrict y to domain=-18.3:1, 
                     col sep = comma] {data/error_sweep/sim10e8/DFR_n0_2_v_9_t_\tvalue.csv};
    \addlegendentryexpanded{$t=\tvalue$};
   }

\end{axis}
\end{tikzpicture}

%% file: figures/tikz_pic_error_comparison.tex
\pgfkeys{/pgf/number format/.cd,1000 sep={}}
\begin{tikzpicture}
  \begin{axis}[scale=0.75,
               legend columns=2,
               xmin = 200,
               xmax = 12000,
               grid = major,
               ymode= log,
               log basis y=10,
               ymin = 2e-8,
               ymax = 1,
               ytick = {1,1e-1,1e-3,1e-5,1e-7},
               legend style={at={(1.27,0.1)},anchor=south, font=\footnotesize},
               mark size=1.5pt,
               xlabel={code length $n$},
               ylabel={DFR},
               line width=0.5pt,
               cycle list name=waterfall-floor-model,
               xticklabel style={
                 /pgf/number format/fixed,
                 /pgf/number format/precision=5
               },
scaled x ticks=false]
   \foreach \tvalue in {10,15,20,...,50} {
     \addplot table [x expr = \thisrow{r}*2,
                     y expr= \thisrow{dfr},
                     restrict y to domain=-18.1:1, 
                     col sep = comma] {data/error_sweep/model/DFR_n0_2_v_9_t_\tvalue.csv};
     \addlegendentryexpanded{};
     \addplot table [x expr = \thisrow{r}*2,
                     y expr = \thisrow{dfr}, 
                     restrict y to domain=-18.3:1, 
                     col sep = comma] {data/error_sweep_fixed/DFR_n0_2_v_9_t_\tvalue.csv};
    \addlegendentryexpanded{$t=\tvalue$};
   }

\end{axis}
\end{tikzpicture}

%% file: figures/tab_BIKE_comparison.tex
\begin{center}
\resizebox{\columnwidth}{!}{
\begin{tabular}{c||ccc|c|c}
\toprule
\textbf{NIST}      &
\multicolumn{3}{c|}{{\bf rate $\frac{1}{2}$ QC-MDPC from}~\cite{BIKE} }  &
\multicolumn{2}{c}{\bf No. of discrepancies: $\weight{\bar{d}^{(2)}}$, with}\\
\textbf{cat.} & $\mathbf{r}$ & $\mathbf{v}$ & $\mathbf{t}$ &
$\mathtt{th}^{(1)}$, $\mathtt{th}^{(2)}$ from~\cite{BIKE}&
{\bf our} $\mathtt{th}^{(1)}$, $\mathtt{th}^{(2)}$ \\
\midrule
$1$ & $12,323$ &  $71$ & $134$ & $17.22$ & $6.39 \times 10^{-2}$\\
$3$ & $24,659$ & $103$ & $199$ & $33.65$ & $7.24 \times 10^{-4}$\\
$5$ & $40,973$ & $137$ & $264$ & $29.92$ & $8.69 \times 10^{-6}$\\
\bottomrule
\end{tabular}
} 
\end{center}

%% file: figures/tab_dfr_leda.tex
\renewcommand{\arraystretch}{1.2}
\begin{tabular}{cccc|c|c}
\toprule
\multicolumn{4}{c}{\textbf{LEDAcrypt parameters}} & \textbf{Fixed} & \textbf{Our variable} \\
\multirow{1}{*}{$\mathbf{n}$} & \multirow{1}{*}{$\mathbf{\frac{k}{n}}$} & \multirow{1}{*}{$\mathbf{v}$} & \multirow{1}{*}{$\mathbf{t}$} & \textbf{$\mathtt{th}^{(1)}$, $\mathtt{th}^{(2)}$}~\cite{extended} & \textbf{ $\mathtt{th}^{(1)}$, $\mathtt{th}^{(2)}$} \\
\midrule
$46742$ &  $1/2$ & $71$ & $130$ &  $2^{-140}$ & $2^{-150}$\\
$48201$ &  $2/3$ & $79$ & $83$  &  $2^{-135}$ & $2^{-143}$\\
$53588$ &  $3/4$ & $83$ & $66$  &  $2^{-131}$ & $2^{-146}$\\
\bottomrule
\end{tabular}

\renewcommand{\arraystretch}{1}